\PassOptionsToPackage{hidelinks}{hyperref}
\PassOptionsToPackage{hyphens}{url}
\documentclass[USenglish,oneside,twocolumn]{article}

\usepackage[utf8]{inputenc}
\usepackage[big]{dgruyter_NEW}

\usepackage{etoolbox}
\makeatletter
\patchcmd{\section}
  {\leavevmode\vrule\@width\z@\@height\dimexpr\topskip+6.5\p@\relax\@depth\z@}{}{}{}
\makeatother
\usepackage[compact]{titlesec}
\titlespacing*{\section}{0pt}{*1}{*1}
\titlespacing*{\subsection}{0pt}{*1}{*1}
\titlespacing*{\subsubsection}{0pt}{*1}{*1}

\cclogo{\includegraphics{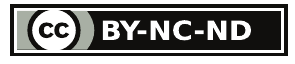}}

\usepackage{amsmath}
\usepackage{mathtools}
\usepackage[capitalize]{cleveref}
\usepackage[acronym]{glossaries}
\usepackage{url}
\usepackage[en-US]{datetime2}
\usepackage{listings}
\usepackage{booktabs} 
\usepackage{tabularx}
\newcolumntype{Y}{>{\raggedleft\arraybackslash}X}
\usepackage{bytefield}
\usepackage{csquotes}
\usepackage{graphicx}
\usepackage{amssymb}
\usepackage{pifont}
\usepackage{multirow}
\usepackage[inline]{enumitem}
\setlist{noitemsep}
\setlist[enumerate,1]{label = (\arabic*),
                      ref   = \arabic*}
\usepackage{caption}
\usepackage{subcaption}
\usepackage{siunitx}
\usepackage{xcolor}
\definecolor{myblue}{HTML}{4C72B0}
\definecolor{myred}{HTML}{C44E52}
\definecolor{mygreen}{HTML}{55A868}
\definecolor{mybluetext}{HTML}{001C7F}
\definecolor{myredtext}{HTML}{8C0800}
\definecolor{mygreentext}{HTML}{12711C}

\definecolor{highlight}{RGB}{192,80,77}
\crefname{lstlisting}{Listing}{Listings}
\Crefname{lstlisting}{Listing}{Listings}
\crefname{section}{\S}{\S\S}
\Crefname{section}{\S}{\S\S}
\crefname{table}{Tab.}{Tabs.}
\Crefname{table}{Table}{Tables}
\crefname{figure}{Fig.}{Figs.}
\Crefname{figure}{Fig.}{Figs.}

\newcommand{\cmark}{\ding{51}}%
\newcommand{\xmark}{\ding{55}}%

\newcommand{\ie}{i.e.}

\newcommand{\eg}{e.g.}
\newcommand{\Eg}{E.g.}

\renewcommand{\paragraph}[1]{{\sffamily\bfseries\mathversion{bold}\normalsize#1.} }

\usepackage[backend=biber, maxcitenames=3, maxbibnames=99, sortcites]{biblatex}

\defbibheading{bibliography}[References]{%
  \section*{#1}%
}

\glsdisablehyper

\newacronym{OF}{OF}{offline finding}
\newacronym{SIP}{SIP}{system integrity protection}
\newacronym{AMFI}{AMFI}{Apple mobile file integrity}
\newacronym{BLE}{BLE}{Bluetooth Low Energy}
\newacronym[longplural={operating systems}, shortplural={OSs}]{OS}{OS}{operating system}
\newacronym{EC}{EC}{elliptic curve}
\newacronym{ECC}{ECC}{elliptic curve cryptography}
\newacronym{ECIES}{ECIES}{Elliptic Curve Integrated Encryption Scheme}
\newacronym{ECDH}{ECDH}{Elliptic Curve Diffie-Hellmann}
\newacronym{IV}{IV}{initialization vector}
\newacronym{AES-GCM}{AES-GCM}{Advanced Encryption Standard in Galois/Counter Mode}
\newacronym{AWDL}{AWDL}{Apple Wireless Direct Link}
\newacronym{DoS}{DoS}{denial-of-service}
\newacronym{MitM}{MitM}{machine-in-the-middle}
\newacronym{PWS}{PWS}{Wi-Fi Password Sharing}
\newacronym{mDNS}{mDNS}{multicast DNS}
\newacronym{ECDSA}{ECDSA}{Elliptic Curve Digital Signature Algorithm}
\newacronym{CA}{CA}{Certificate Authority}
\newacronym{SEP}{SEP}{Secure Enclave Processor}
\newacronym{TEE}{TEE}{trusted execution environment}
\newacronym{FIPS}{FIPS}{U.S. Federal Information Processing Standards}
\newacronym{LOWESS}{LOWESS}{locally weighted scatterplot smoothing}
\newacronym{TLS}{TLS}{Transport Layer Security}

\begin{document}

  \author*[1]{Alexander Heinrich}

  \author*[2]{Milan Stute}

  \author[3]{Tim Kornhuber}

  \author[4]{Matthias Hollick}

  \affil[1]{Secure Mobile Networking Lab, Technical University of Darmstadt, Germany, E-mail: aheinrich@seemoo.tu-darmstadt.de}

  \affil[2]{Secure Mobile Networking Lab, Technical University of Darmstadt, Germany,\newline E-mail: mstute@seemoo.tu-darmstadt.de}

  \affil[3]{Secure Mobile Networking Lab, Technical University of Darmstadt, Germany,\newline E-mail: tkornhuber@seemoo.tu-darmstadt.de}

  \affil[4]{Secure Mobile Networking Lab, Technical University of Darmstadt, Germany,\newline E-mail: mhollick@seemoo.tu-darmstadt.de}

  \title{\huge Who Can \emph{Find My} Devices? \\ Security and Privacy of Apple's Crowd-Sourced Bluetooth Location Tracking System}

  \runningtitle{Who Can \emph{Find My} Devices? Security and Privacy of Apple's Crowd-Sourced Bluetooth Location Tracking System}


\begin{abstract}
{
Overnight, Apple has turned its hundreds-of-million-device ecosystem into the world's largest crowd-sourced location tracking network called \gls{OF}. \gls{OF} leverages online finder devices to detect the presence of missing offline devices using Bluetooth and report an approximate location back to the owner via the Internet.
While \gls{OF} is not the first system of its kind, it is the first to commit to strong privacy goals. In particular, \gls{OF} aims to ensure finder anonymity, untrackability of owner devices, and confidentiality of location reports.
This paper presents the first comprehensive security and privacy analysis of \gls{OF}.
To this end, we recover the specifications of the closed-source \gls{OF} protocols by means of reverse engineering. 
We experimentally show that unauthorized access to the location reports allows for accurate device tracking and retrieving a user's top locations with an error in the order of 10 meters in urban areas.
While we find that \gls{OF}'s design achieves its privacy goals, we discover two distinct design and implementation flaws that can lead to a location correlation attack and unauthorized access to the location history of the past seven days, which could deanonymize users. 
Apple has partially addressed the issues following our responsible disclosure. Finally, we make our research artifacts publicly available.
}
\end{abstract}


  \journalname{Proceedings on Privacy Enhancing Technologies}
  \DOI{Editor to enter DOI}
  \startpage{1}
  \received{..}
  \revised{..}
  \accepted{..}

  \journalyear{..}
  \journalvolume{..}
  \journalissue{..}

\maketitle


\section{Introduction}
\label{sec:introduction}
\glsresetall

In 2019, Apple introduced \emph{\gls{OF}}, a proprietary crowd-sourced location tracking system for offline devices.
The basic idea behind \gls{OF} is that so-called \emph{finder} devices can detect the presence of other \emph{lost} offline devices using \gls{BLE} and use their Internet connection to report an approximate location back to the \emph{owner}.
Apple's \gls{OF} network consists of \enquote{hundreds of millions} of devices~\cite{Apple:PlatformSecurity_Spring2020}, making it the currently largest crowd-sourced location tracking system in existence. We expect the network to grow as \gls{OF} will officially support the tracking of non-Apple devices in the future~\cite{AppleFindMyNetworkSpecification}.
Regardless of its size, the system has sparked considerable interest and discussion within the broader tech and security communities~\cite{Wired2019,Green2019} as Apple makes strong security and privacy claims supported by new cryptographic primitives that other commercial systems are lacking~\cite{Weller_2020}. In particular, Apple claims that it cannot access location reports, finder identities are not revealed, and \gls{BLE} advertisements cannot be used to track devices~\cite{Apple:BlackHat2019}.
Apple has yet to provide ample proof for their claims as, until today, only selected components have been publicized~\cite{Apple:PlatformSecurity_Spring2020,AppleFindMyNetworkSpecification,Apple:BlackHat2019}.

\paragraph{Contribution}
This paper challenges Apple's security and privacy claims and examines the system design and implementation for vulnerabilities.
To this end, we first analyze the involved \gls{OF} system components on macOS and iOS using reverse engineering and present the proprietary protocols involved during \emph{losing}, \emph{searching}, and \emph{finding} devices.
In short, devices of one owner agree on a set of so-called rolling public--private key pairs.
Devices without an Internet connection, \ie, without cellular or Wi-Fi connectivity,  emit \gls{BLE} advertisements that encode one of the rolling public keys. Finder devices overhearing the advertisements encrypt their current location under the rolling public key and send the location report to a central Apple-run server. When searching for a lost device, another owner device queries the central server for location reports with a set of known rolling public keys of the lost device. The owner can decrypt the reports using the corresponding private key and retrieve the location.

Based on our analysis, we assess the security and privacy of the \gls{OF} system. 
We find that the overall design achieves Apple's specific goals. 
However, we discovered two distinct design and implementation vulnerabilities that seem to be outside of Apple's threat model but can have severe consequences for the users.
First, the \gls{OF} design allows Apple to correlate different owners' locations if their locations are reported by the same finder, effectively allowing Apple to construct a social graph.
Second, malicious macOS applications can retrieve and decrypt the \gls{OF} location reports of the last seven days for all its users and for \emph{all} of their devices as cached rolling advertisement keys are stored on the file system in cleartext. 
We demonstrate that the latter vulnerability is exploitable and verify that the accuracy of the retrieved reports---in fact---allows the attacker to locate and identify their victim with high accuracy.
We have shared our findings with Apple via responsible disclosure, who have meanwhile fixed one issue via an OS update (CVE-2020-9986, cf.~\emph{Responsible Disclosure} section for details).
We summarize our key contributions.
\begin{itemize}
	\item We provide a comprehensive specification of the \gls{OF} protocol components for losing, searching, and finding devices. Our PoC implementation allows for tracking non-Apple devices via Apple's \gls{OF} network.
	\item We experimentally evaluate the accuracy of real-world location reports for different forms of mobility (by car, train, and on foot). We show that
	(1)	a walking user's path can be tracked with a mean error of less than \SI{30}{m} in a metropolitan area and
	(2) the top locations of a user such as home and workplace can be inferred reliably and precisely (error in the order of \SI{10}{m}) from a one-week location trace.
	\item We discover a design flaw in \gls{OF} that lets Apple correlate the location of multiple owners if the same finder submits the reports. This would jeopardize location privacy for all other owners if only a single location became known.
	\item We discover that a local application on macOS can effectively circumvent Apple's restrictive location API~\cite{Apple:CoreLocation} and access the user's location history without their consent, allowing for device tracking and user identification.
	\item We open-source our PoC implementation and experimental data~(cf.~\emph{Availability} section).
\end{itemize}

\paragraph{Outline}
The remainder of this paper is structured as follows.
\Cref{sec:background} and \cref{sec:overview} provide background information about \gls{OF} and the involved technology.
\Cref{sec:adversary_model} outlines our adversary model.
\Cref{sec:methodology} summarizes our reverse engineering methodology.
\Cref{sec:protocol} describes the \gls{OF} protocols and components in detail.
\Cref{sec:evaluation} evaluates the accuracy of \gls{OF} location reports.
\Cref{sec:security_analysis} assesses the security and privacy of Apple's \gls{OF} design and implementation.
\Cref{sec:vuln1_correlate} and \cref{sec:vuln2_location_access} report two discovered vulnerabilities and propose our mitigations.
\Cref{sec:relatedwork} reviews related work.
Finally, \cref{sec:conclusion} concludes this work.


\section{Background}
\label{sec:background}

This section gives a brief introduction to \gls{BLE} and \gls{ECC} as they are the basic building blocks for \gls{OF}. We then cover relevant Apple platform internals.

\subsection{Bluetooth Low Energy}
\glsreset{BLE}

\gls{BLE}~\cite{Bluetooth5_2:CoreSpec} is designed for small battery-powered devices such as smartwatches and fitness trackers with low data rates.
Devices can broadcast \gls{BLE} advertisements to inform nearby devices about their presence. The maximum \gls{BLE} advertisement payload size is 31 bytes~\cite{Bluetooth5_2:CoreSpec}.
Apple heavily relies on custom \gls{BLE} advertisements to announce their proprietary services such as AirDrop and bootstrap their protocols over Wi-Fi or \gls{AWDL}~\cite{Stute2019,Martin2019,Celosia2020}.
\gls{OF} devices also use \gls{BLE} advertisements to inform nearby finders about their presence~\cite{AppleFindMyNetworkSpecification}.

\subsection{Elliptic Curve Cryptography}
\label{sec:background:ecc}
\glsreset{ECC}

\gls{OF} employs \gls{ECC} for encrypting location reports. \gls{ECC} is a public-key encryption scheme that uses operations on \gls{EC} over finite fields.
An \gls{EC} is a curve over a finite field that contains a known generator (or base point) $G$. 
A private key in \gls{ECC} is a random number in the finite field of the used curve.
The public key is the result of the point multiplication of the generator $G$ with the private key. The result is an X--Y coordinate on the curve.
The NIST P-224 curve~\cite{NISTDSS}, which is used by \gls{OF}~\cite{AppleFindMyNetworkSpecification}, provides a security level of 112 bit.

\subsection{Apple Platform Internals}

We briefly introduce the terms keychain and iCloud as they are relevant for Apple's \gls{OF} implementation.

\paragraph{Keychain}
All Apple \glspl{OS} use a keychain as a database to store secrets such as passwords, keys, and trusted \gls{TLS} root certificates.
The keychain is used by system services such as AirDrop~\cite{Stute2019} and third-party applications to store login information, tokens, and other secrets.
Every keychain item may contain a \emph{keychain access group}. This group is used to identify which application can access which keychain items. Access policies are implemented via \emph{entitlement} files embedded into signed application binaries. A system process prevents the execution of processes with unauthorized entitlements, \eg, a third-party application trying to access a system-owned keychain item.
This security mechanism can be disabled on jailbroken iOS devices or by deactivating macOS \gls{SIP}, which helps extracting keys and secrets used by Apple's system services.

\paragraph{iCloud}
iCloud is an umbrella term for all Apple services handling online data storage and synchronization via Apple's servers.
All \emph{owner} devices signed in to the same Apple account can synchronize themselves via iCloud.
\gls{OF} uses the iCloud keychain to share rolling advertisement keys across all owner devices. The synchronization is required to retrieve and decrypt the location reports from potential finders on any of the owner devices~\cite{Apple:BlackHat2019,Apple:PlatformSecurity_Spring2020}.

\section{Apple Offline Finding Overview}
\label{sec:overview}

Apple introduced \gls{OF} in~2019 for iOS~13, macOS~10.15, and watchOS~6~\cite{AppleKeynoteWWDC2019}. 
\gls{OF} enables locating Apple devices without an Internet connection and promises to operate in a privacy-preserving manner.
In 2020, Apple announced to support third-party \gls{BLE}-enabled devices to be tracked by the \gls{OF} network~\cite{AppleKeynoteWWDC2020} and released a protocol specification for their integration~\cite{AppleFindMyNetworkSpecification}. We found that this public specification is incomplete concerning the overall \gls{OF} system.
Within this paper, we focus on our recovered specification that was partly validated by the accessory specification~\cite{AppleFindMyNetworkSpecification}.

\begin{figure}
  \includegraphics[width=\linewidth]{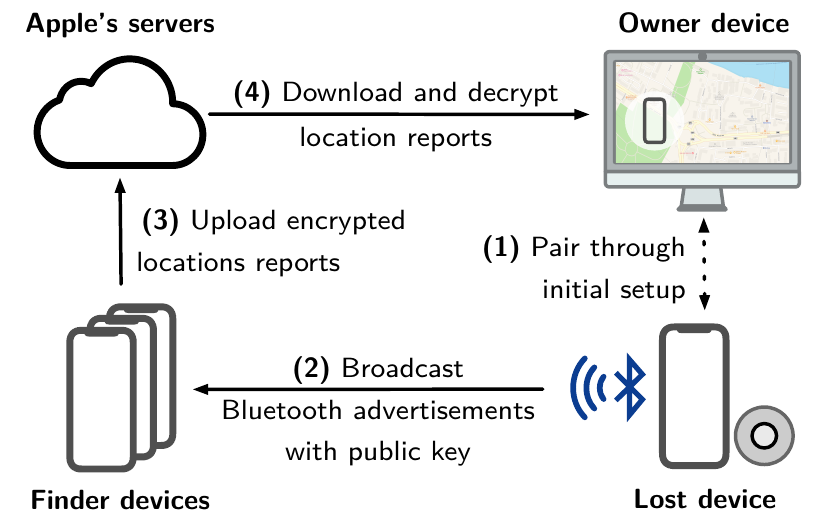}
  \caption{Simplified offline finding (OF) workflow.}
  \label{fig:of_overview}
\end{figure}

In the following, we give a brief overview of how \gls{OF} works and introduce the different roles of devices. \Cref{fig:of_overview} depicts the interplay of the roles and protocols involved in \gls{OF}.
In particular, \gls{OF} involves (1) initial pairing of owner devices, (2) broadcasting \gls{BLE} advertisements that contain a rolling public key, (3) uploading encrypted location reports to Apple's servers, and (4) retrieving the location reports on owner devices.
The terminology of the roles below has been derived from the official documentation~\cite{AppleFindMyNetworkSpecification}.

\paragraph{Owner devices}
	Owner devices share a common Apple ID and can use the \emph{Find My} application on macOS and iOS to search for any devices of the same owner.

\paragraph{Lost devices}
	Devices that determine to be in a lost state start sending out \gls{BLE} advertisements with a public key to be discovered by finder devices.
	Apple devices are considered to be lost when they lose Internet connectivity.
	Third-party accessories~\cite{AppleFindMyNetworkSpecification} are small battery-powered devices that can be attached to a personal item and are set up through an owner device. Accessories determine to be \emph{lost} when they lose their \gls{BLE} connection to the owner device.

\paragraph{Finder devices}
	Finder devices form the core of the \gls{OF} network. As of 2020, only iPhones and iPads with a GPS module are offering finder capabilities. Finder devices can discover lost devices and accessories by scanning for \gls{BLE} advertisements. Upon receiving an \gls{OF} advertisement, a finder creates an end-to-end encrypted location report that includes its current location and sends it to Apple's servers.

\paragraph{Apple's servers}
	Apple's servers store \gls{OF} location reports submitted by finder devices. Owner devices can fetch those reports and decrypt them locally.


\section{Adversary Model}
\label{sec:adversary_model}

\begin{table*}
	\caption{Adversary models considered throughout and guiding this paper. We assume that neither adversary has direct access to cryptographic secrets or can break cryptographic primitives.}
	\label{tab:adversary-model}
	\small
	\begin{tabularx}{\linewidth}{lXXX}
		\toprule
		\textbf{Model} & \textbf{Assumptions} & \textbf{Goals} & \textbf{Capabilities} \\
		\midrule

		Local application \textbf{(A1)} &
		\begin{enumerate*}
			\item User-installed application on lost/owner devices that is either reviewed or notarized.
			\item Zero-permission.
			\item No privilege escalation exploits.
		\end{enumerate*} &
		\begin{enumerate*}
			\item Determine location of \gls{OF} devices without asking for permission.
			\item Track user by accessing current or historic location data.
		\end{enumerate*} &
		\begin{enumerate*}
			\item Communicate with any server over the Internet.
			\item Read/write files that are accessible by the user and not restricted through sandboxing.
		\end{enumerate*} \\

		Proximity-based \textbf{(A2)} &
		\begin{enumerate*}
			\item In \gls{BLE} communication range of \gls{OF} device.
			\item Control one or more \gls{BLE} transceivers to cover a larger area.
		\end{enumerate*} &
		\begin{enumerate*}
			\item Access location of lost devices or personally linkable data.
			\item Track lost devices in larger areas (\eg, shopping center or airport).
			\item \gls{DoS} against \gls{OF} service.
		\end{enumerate*} &
		\begin{enumerate*}
			\item Track devices based on advertisement content.
			\item Record and replay advertisements at different locations.
			\item Fabricate new advertisements.
		\end{enumerate*} \\

		Network-based \textbf{(A3)} &
		\begin{enumerate*}
			\item \gls{MitM} position between Apple and \gls{OF} devices.
			\item Cannot break \gls{TLS}.
		\end{enumerate*} &
		\begin{enumerate*}
			\item Access location of reported lost devices.
			\item Identify reported devices.
			\item Identify lost devices.
		\end{enumerate*} &
		\begin{enumerate*}
			\item Redirect traffic to a different host.
			\item Read, intercept, redirect, or modify traffic.
		\end{enumerate*} \\

		Service operator \textbf{(A4)} &
		\begin{enumerate*}
			\item Apple as the service provider.
			\item Controls the \gls{OF} server infrastructure.
		\end{enumerate*} &
		\begin{enumerate*}
			\item Locate individuals and their lost devices.
			\item Correlate locations to create a social graph.
		\end{enumerate*} &
		\begin{enumerate*}
			\item Access to all encrypted \gls{OF} reports and their metadata.
			\item Add, remove, or modify reports.
		\end{enumerate*} \\

		\bottomrule
	\end{tabularx}
\end{table*}

\gls{OF} exposes several interfaces that might be targeted by attackers.
In this section, we identify these potentially vulnerable interfaces and devise a comprehensive adversary model that will guide the rest of this paper.
We first detail the four sub-models, summarized in \cref{tab:adversary-model}, and we specify them by their assumptions, goals, and capabilities following~\cite{doRoleAdversaryModel2019}. Then, we motivate the subsequent analysis of \gls{OF} protocols and components based on these models.

First of all, we consider adversaries on either of \gls{OF}'s communication channels (cf.~(2)--(4) in \cref{fig:of_overview}). In particular, a proximity-based adversary has access to \gls{BLE} advertisements~\textbf{(A2)}, and a network-based adversary can modify traffic between \gls{OF} devices and Apple's servers~\textbf{(A3)}.
Also, we consider a zero-permission application running with user privileges on an owner/lost device that wants to infer the user's current location. The application may be distributed inside or outside\footnote{On macOS, applications can be distributed outside of Apple's app store. Those applications are not reviewed~\cite{Apple:AppReview}. However, Apple recommends submitting those application to their notarization service, which checks application for malicious code~\cite{Apple:Notarization}. macOS displays a warning and asks for user consent before executing non-notarized applications.} of Apple's official app stores~\textbf{(A1)}.
Finally, we also consider Apple as the service operator as an adversary that has access to all encrypted location reports and might try to infer any information based on the report metadata such as submission times and finder identifiers~\textbf{(A4)}.
Note that Apple uses its iCloud keychain service for initial device pairing and key synchronization (cf.~(1) in~\cref{fig:of_overview}). Apple provides detailed information about its keychain~\cite{Apple:PlatformSecurity_Spring2020}, which appears to withstand professional forensics analyses~\cite{Elcomsoft:iOSKeychain}. Therefore, we assume that the pairing process is secure throughout this paper.

To conduct a security and privacy analysis based on these models, we need to understand \gls{OF} in detail. To this end, we reverse engineer the protocols involved in loosing, finding, and searching devices (cf.~(2)--(4) in \cref{fig:of_overview}) in \cref{sec:protocol}.
Based on our understanding of \gls{OF}, we conduct a security and privacy analysis of the \gls{BLE} communication \textbf{(A2)}, the server communication \textbf{(A3)}, and storage of encrypted reports and cryptographic keys~\textbf{(A1/A4)} in \cref{sec:security_analysis}. 


\section{Methodology}
\label{sec:methodology}

Our analysis of \gls{OF} required a comprehensive understanding of the implemented protocols by Apple. Our methodology follows previous works analyzing the Apple ecosystem~\cite{Stute2019,Stute2020,Celosia2020,Martin2019,Stute2021}, while providing new insights into the reverse engineering process.
We started this research with the beta releases of macOS~10.15 and iOS~13, the first Apple \glspl{OS} to support \gls{OF}. 
During that time, no official documentation from Apple was available regarding the \gls{OF} design or implementation. Therefore, we used reverse engineering tools such as system log analysis, static binary analysis, and network traffic analysis.
In addition, we implemented an \gls{OF} prototype to validate our findings.
Some of our findings, such as the \gls{BLE} advertisement format and cryptographic primitives, were later confirmed by Apple's specification for third-party accessories~\cite{AppleFindMyNetworkSpecification}.

\paragraph{System Logging}
To get a first overview of \gls{OS} internals, we used the system logging facility on macOS. It aggregates applications and kernel events, and can access the same events from a USB-attached iOS device. We can filter logs by process or keyword and adjust the log level for more verbose output. By using a special configuration profile~\cite{garside:ShowPrivateLog}, macOS will show logs that are normally redacted. On iOS, this option is only available with a jailbreak~\cite{arbuckleUnredactprivateosLogs}. 

\paragraph{Binary analysis}
We use binary analysis to understand the closed-source \gls{OF} protocols. Many Apple binaries have been written in {Objective-C}, which uses message dispatch to resolve methods at runtime. Therefore, {Objective-C} binaries include method and instance variable names as part of the dispatch table. This simplifies identifying interesting code paths and segments, \eg, those responsible for parsing \gls{BLE} packets.
Unfortunately, most \gls{OF} code is written in the newer Swift programming language. Swift methods are statically called by their program address and, therefore, do not require an entry in the symbol table, \ie, the symbol names may be stripped by the compiler. Additionally, the Swift compiler adds several checks to achieve type safety, which clutters the compiled code and makes it hard to follow the program logic. However, dynamically linked frameworks and libraries must keep function names in the symbol table, facilitating the identification of interesting code segments.
Furthermore, dynamic analysis methods aid in understanding the control flow and access function parameters at runtime. By hooking functions with a dynamic instrumentation tool such as Frida~\cite{FRIDA}, we can, \eg, access cryptographic keys used by system processes as shown in~\cite{Stute2021}.

\paragraph{Network analysis}
We can identify a service's protocols by monitoring network interfaces, which helps understand the information exchange with external parties. \gls{OF} uses two protocols: \gls{BLE} for advertisements and HTTPS for server communication.
To understand the embedded custom protocols and payloads, we rely on two sets of tools.
For \gls{BLE}, we use BTLEmap~\cite{Heinrich:BTLEmap} to capture all \gls{BLE} advertisements. As we already know the basic frame format of Apple's custom advertisements from related work~\cite{Celosia2020,Martin2019}, we were able to identify \gls{OF} as a new subtype.
HTTPS proxies such as~\cite{Tran:Proxyman} decrypt HTTPS sessions by masquerading as both HTTP client and server and using self-signed \gls{TLS} certificates.
To access \gls{OF}-related traffic, we disabled \emph{certificate pinning}, which \gls{OF} clients use for all server communication. 


\section{Apple Offline Finding in Detail}
\label{sec:protocol}

This section describes and discusses the technical details of Apple's \gls{OF} system.
In reference to \cref{fig:of_overview}, we
\begin{enumerate*}
	\item explain the involved cryptography and the key exchange during initial device pairing, and then explain the protocols implementing
	\item \emph{losing},
	\item \emph{finding},
	\item \emph{searching} for devices.
\end{enumerate*}

In short, devices and accessories in lost mode send out \gls{BLE} advertisements containing a public key. Finder devices receive them, encrypt their location by using the public key, and upload a report to Apple's servers. This results in an end-to-end encrypted location report that cannot be read by Apple or any other third-party that does not have access to the owner's private keys.
In the following, we explain the cryptography in use, the protocols involved in losing, searching, and finding devices, as well as a brief description of the system's implementation on iOS and macOS.

\subsection{Cryptography}
\label{sec:Cryptography}

\gls{OF} employs \gls{ECC}~\cite{AppleFindMyNetworkSpecification}. In the following, we explain the key generation and derivation mechanisms and the cryptographic algorithms used for encryption and decryption. 

\paragraph{Master Beacon and Advertisement Keys}
Initially, each owner device generates a private--public key pair $(d_0, p_0)$ on the NIST P-224 curve and a 32-byte symmetric key $SK_0$ that together form the \emph{master beacon key}.
Those keys are never sent out via \gls{BLE} and are used to derive the rolling advertisement keys included in the \gls{BLE} advertisements.

\gls{OF} makes device tracking hard by regularly changing the contents of the \gls{BLE} advertisements.
In particular, \gls{OF} uses the concept of \emph{rolling} keys that can be deterministically derived if one knows the initial input keys $(d_0, p_0)$ and $SK_0$ but are otherwise unlinkable.
\gls{OF} iteratively calculates the \emph{advertisement keys} $(d_i, p_i)$ for $i>0$ as follows using the ANSI X.963 key derivation function (KDF) with SHA-256~\cite{X963} and a generator $G$ of the NIST P-224 curve: 
\begin{align}
	SK_{i} &= \text{KDF}( SK_{i-1}, \text{\enquote{update}}, 32) \label{eq:key_derivation:1} \\
	(u_{i}, v_{i}) &= \text{KDF}(SK_{i}, \text{\enquote{diversify}}, 72) \label{eq:key_derivation:2} \\
	d_{i} &= (d_0 * u_{i}) + v_{i} \label{eq:key_derivation:3} \\
	p_{i} &= d_{i} * G \label{eq:key_derivation:4}
\end{align}
\Cref{eq:key_derivation:1} derives a new symmetric key from the last used symmetric key with 32 bytes length. 
\Cref{eq:key_derivation:2} derives the so-called \enquote{anti-tracking} keys $u_{i}$ and $v_{i}$ from the new symmetric key with a length of 36 bytes each.
Finally, \cref{eq:key_derivation:3,eq:key_derivation:4} create the advertisement key pair via \gls{EC} point multiplication using the anti-tracking keys and the master beacon key $d_0$.

\paragraph{Key Synchronization}
All owner devices need to access the advertisement keys to download and decrypt location reports. Therefore, \gls{OF} synchronizes the master beacon keys via iCloud in a property list file encrypted under \gls{AES-GCM}. The decryption key for the file is stored in the iCloud keychain under the label \enquote{Beacon Store.}

\paragraph{Encryption}
The \gls{BLE} advertisements sent out by a lost device contain an \gls{EC} public key $p_i$. A finder device that receives such an advertisement determines its current location and encrypts the location with $p_i$. \gls{OF} employs \gls{ECIES} that performs an ephemeral \gls{ECDH} key exchange to derive a shared secret and encrypt the report~\cite{rfc-6090-ecc}. In particular, the finder's encryption algorithm works as follows:
\begin{enumerate}
	\item Generate a new ephemeral key $(d',p')$ on the NIST P-224 curve for a received \gls{OF} advertisement.
	\item Perform \gls{ECDH} using the ephemeral private key $d'$ and the advertised public key $p_i$ to generate a shared secret.
	\item Derive a symmetric key with ANSI X.963 KDF on the shared secret with the advertised public key as entropy and SHA-256 as the hash function.
	\item Use the first 16 bytes as the encryption key $e'$.
	\item Use the last 16 bytes as an \gls{IV}.
	\item Encrypt the location report under $e'$ and the \gls{IV} with \gls{AES-GCM}.
\end{enumerate}
The ephemeral public key $p'$ and the authentication tag of \gls{AES-GCM} are part of the uploaded message, as shown in \cref{fig:location_report_binary}. All location reports are identified by an id, which is a SHA-256 hash of $p_i$.

\paragraph{Decryption}
An owner device that retrieves encrypted location reports follows the inverse of the encryption procedure.
First, the owner device selects the proper advertisement keys $(d_i,p_i)$ based on the hashed $p_i$ of the location report.
Second, it performs the \gls{ECDH} key exchange with the finder's ephemeral public key $p'$ and the lost device's private key $d_i$ to compute the symmetric key $e'$ and the IV.
Finally, the owner can use $e'$ and IV to decrypt the location report.

\subsection{Losing}
\label{sec:protocol_losing}

An \gls{OF} device that loses its Internet connection starts emitting \gls{BLE} advertisements. This advertisement consists of the 224 bit (28 bytes) public part\footnote{More precisely, \gls{OF} only advertises the X coordinate of the public key, which has a length of 28 bytes. The Y coordinate is irrelevant for calculating a shared secret via \gls{ECDH}, so the sign bit for the compressed format~\cite{DRLBrown:EllipticCurveCryptography} can be omitted.}
of the advertisement key ($p_i$), but required some engineering effort to fit in a single \gls{BLE} packet.

\begin{table}
\caption{\gls{OF} advertisement format (with zero-indexed bytes).}
\label{tab:BLE_Payload}
	\small
	\begin{tabularx}{\linewidth}{rX}
	\toprule
	Bytes & Content (details cf.~\cite[§~5.1]{AppleFindMyNetworkSpecification}) \\ 
	\midrule
	0--5 & BLE address ($(p_i[0] \mathbin{|} (0b11 \ll 6)) \mathbin{||} p_i[1..5]$) \\
	\midrule
	6 & Payload length in bytes (\lstinline{30}) \\ 
	7 & Advertisement type (\lstinline{0xFF} for manufacturer-specific data) \\ 
	8--9 & Company ID (\lstinline{0x004C}) \\ 
	10 & \gls{OF} type (\lstinline{0x12}) \\ 
	11 & \gls{OF} data length in bytes (\lstinline{25}) \\ 
	12 & Status (\eg, battery level) \\ 
	13--34 & Public key bytes $p_i[6..27]$ \\
	35 & Public key bits $p_i[0] \gg 6$ \\
	36 & Hint (\lstinline{0x00} on iOS reports) \\
	\bottomrule
	\end{tabularx}
\end{table}

\paragraph{Advertisement Packet Format}
Apple had to engineer its way around the fact that one \gls{BLE} advertisement packet may contain at most 37 bytes~\cite[Vol.~6, Part~B, §~2.3.1.3]{Bluetooth5_2:CoreSpec}, of which 6 bytes are reserved for the advertising MAC address, and up to 31 can be used for the payload.
For standard compliance, the custom \gls{OF} advertisements needs to add a 4-byte header for specifying \emph{manufacturer-specific data}, which leaves 27 bytes. Within this space, Apple uses a custom encoding for subtypes used by other wireless services such as AirDrop~\cite{Celosia2020}), which leaves 25 bytes for \gls{OF} data.
To fit the 28-byte advertisement key in one packet, Apple re-purposes the random address field to encode the key's first 6 bytes.
However, there is one caveat: the \gls{BLE} standard requires that the first two bits of a random address be set to $0b11$.
\gls{OF} stores the first two bits of the advertisement key together with the 24 remaining bytes in the payload to solve the problem.
We depict the complete \gls{BLE} advertisement packet format in \cref{tab:BLE_Payload}. Apple confirmed the reverse-engineered specification later~\cite{AppleFindMyNetworkSpecification}. 

\paragraph{Advertising Interval}
The same key is emitted during a window of 15 minutes, after which the next key $p_{i+1}$ is used.
During a window, \gls{OF}-enabled iOS and macOS devices emit one \gls{BLE} advertisement every two seconds when they lose Internet connectivity.

\subsection{Finding}
\label{sec:protocol_finding}

All finder devices regularly scan for \gls{OF} advertisements. When the finder receives a packet in the \gls{OF} advertisement format, it generates and uploads an encrypted location report to Apple's servers.

\begin{figure}
	\includegraphics[width=\linewidth]{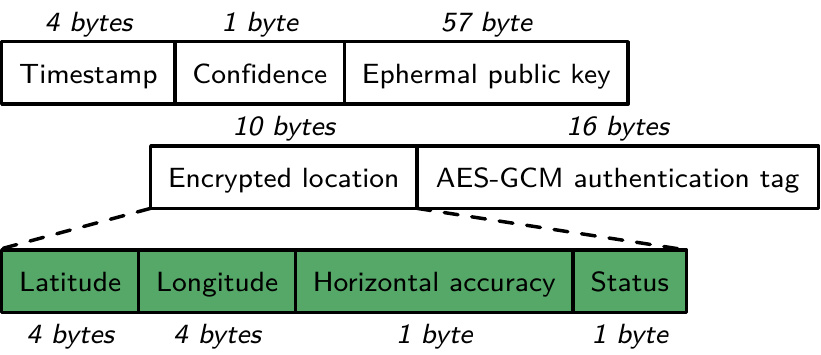}
	\caption{Binary format of a location report.}
	\label{fig:location_report_binary}
\end{figure}

\paragraph{Generating Reports} 
The finder parses the public key from the advertisement. Then, it determines its current geolocation and creates a message that includes location, accuracy,\footnote{We assume that the accuracy value is encoded in metric meters as it matches the experimentally determined positioning error of the coordinates in the location reports, as we show in~\cref{sec:evaluation}.} and status information (cf.~\textcolor{mygreentext}{green} fields in \cref{fig:location_report_binary}). The message is then encrypted using the algorithm described in~\cref{sec:Cryptography}.
Finally, the finder creates a complete location report, including the current timestamp (in seconds since January 1, 2001), the ephemeral public key $d'$, the encrypted message, and the \gls{AES-GCM} authentication tag as shown in \cref{fig:location_report_binary}.

\paragraph{Uploading Reports}
Finder devices accumulate reports over time and upload them in batches regularly, possibly reducing energy and bandwidth consumption. During the evaluation with our test devices, we discovered that the median time from generating to uploading a location report is \SI{26}{min}. We include the delay distribution in \cref{sec:reporting-delay}. The delay can increase to several hours if the finder device is in a low power mode~\cite{AppleLowPowerMode}. A finder limits the number of uploaded reports for the same advertisement key to four, most likely to prevent excess traffic on Apple's servers.
The upload is implemented as an HTTPS POST request to \url{https://gateway.icloud.com/acsnservice/submit}. Every request is authenticated to ensure that only genuine Apple devices can upload requests. \Cref{tab:submit_reports_headers} shows the request header containing a device identity certificate, the signing CA's certificate, and an \gls{ECDSA} signature over the request body. The certificates are stored in the device's keychain. However, the private key used for signing is stored in the \gls{SEP}, Apple's implementation of a \gls{TEE}~\cite{Apple:PlatformSecurity_Spring2020}.
The \gls{SEP} prohibits the extraction of the signing key but provides an interface for signing requests.
We assume that the finder authentication serves as a form of remote attestation. However, we were unable to verify this assumption due to the obfuscated code.
The HTTPS request body is prefixed with a fixed header (\lstinline{0x0F8AE0}) and one byte specifying the number of included reports. This limits the number of reports in a single request to 255. Each report consists the ID ($\text{SHA-256}(p_i)$) followed by the 88-byte location report shown in \cref{fig:location_report_binary}.

\begin{table}
\caption{HTTP request headers for uploading location reports.}
\label{tab:submit_reports_headers}

	\small
	\begin{tabularx}{\linewidth}{lX}
	\toprule
	\textbf{Request Header} & \textbf{Value} \\ 
	\midrule
	X-Apple-Sign1 & Device identity certificate (base64) \\ 
	X-Apple-Sign2 & SHA-256 hash of the signing CA (base64) \\ 
	X-Apple-Sign3 & Device ECDSA signature (ASN.1) \\ 
	X-Apple-I-TimeZone & Client's time zone (\eg, GMT+9) \\
	X-Apple-I-ClientTime & Client's time (Unix) \\ 
	User-Agent & \enquote{searchpartyd/1 <iPhoneModel>/<OSVersion>} \\ 
	\bottomrule
	\end{tabularx}
\end{table}

\subsection{Searching}
\label{sec:protocol_searching}

An owner requests reported location from Apple's servers when searching for a lost device.
As the advertisement keys are synchronized across all of the owner's devices, the owner can use any of their other devices with Apple's \emph{Find My} app to download and decrypt the location reports.
In short, the owner device fetches location reports from Apple's servers by sending a list of the most recent public advertisement keys of the lost device.

\paragraph{Downloading Reports}
Similar to uploading~(cf.~\cref{sec:protocol_searching}), downloading is implemented as an HTTPS POST request to \url{https://gateway.icloud.com/acsnservice/fetch}.
We show the headers in \cref{tab:fetch_reports_headers} and a truncated example body in~\cref{appendix:http}.
The user authenticates with Apple's servers using their Apple account in two steps. First, HTTP basic authentication~\cite{RFC7617} is performed with a unique identifier of the user's Apple ID\footnote{This numerical identifier is unique to each Apple account and does not change even if the user changes their primary email address.} and a \emph{search-party-token} that is device-specific and changes at irregular intervals (in the order of weeks).
Second, several headers with so-called \enquote{anisette data} are included. Anisette data are short-lived tokens valid for \SI{30}{s} and allow omitting two-factor authentication from a previously authenticated system~\cite{Elcomsoft:iCloudAuthentication}.

\begin{table}
\caption{HTTP request headers for downloading location reports.}
\label{tab:fetch_reports_headers}

	\small
	\begin{tabularx}{\linewidth}{l@{}X}
	\toprule
	\textbf{Request Header} & \textbf{Value} \\ 
	\midrule

	Authorization & Base64 encoded basic authentication \\ 
	X-Apple-I-MD & Anisette data \\ 
	X-Apple-I-MD-RINFO & Anisette data \\ 
	X-Apple-I-MD-M & Anisette data \\ 
	X-Apple-I-TimeZone & Client's time zone \\
	X-Apple-I-ClientTime & Client's time (ISO 8601) \\ 
	X-BA-CLIENT-TIMESTAMP & ~~Client's time (Unix) \\ 
	User-Agent & \enquote{searchpartyd/1 <iPhoneModel>/<OSVersion>} \\ 

	\bottomrule

	\end{tabularx}
\end{table}

\paragraph{Decrypting Reports}
The response to the download request contains a list of finder location reports (cf.~\cref{fig:location_report_binary}) and metadata such as the hashed public advertisement key and the time when the report was uploaded.
We show a truncated example of the response body in \cref{appendix:http}.
Using the respective private advertisement keys $d_i$, the owner device can then decrypt the received location reports.
Apple's \emph{Find My} application combines a subset of the reports to display the most recent location of the lost device on a map.
According to Apple, multiple reports are combined to get a more accurate location~\cite[p.~104]{Apple:PlatformSecurity_Spring2020}. While we did not reconstruct Apple's algorithm, we show in \cref{sec:evaluation} that the downloaded location reports are sufficient to not only determine the most recent location but to even precisely reconstruct and trace the movement of a lost device.

\subsection{System Implementation}

\begin{figure}
	\includegraphics[width=\linewidth]{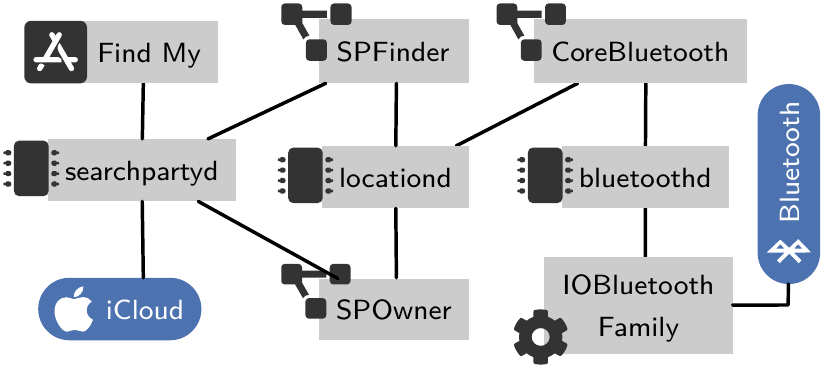}
	\caption[Simplified view of components and their interactions]{Simplified view of components and their interactions such as apps (\includegraphics[height=\fontcharht\font`\B]{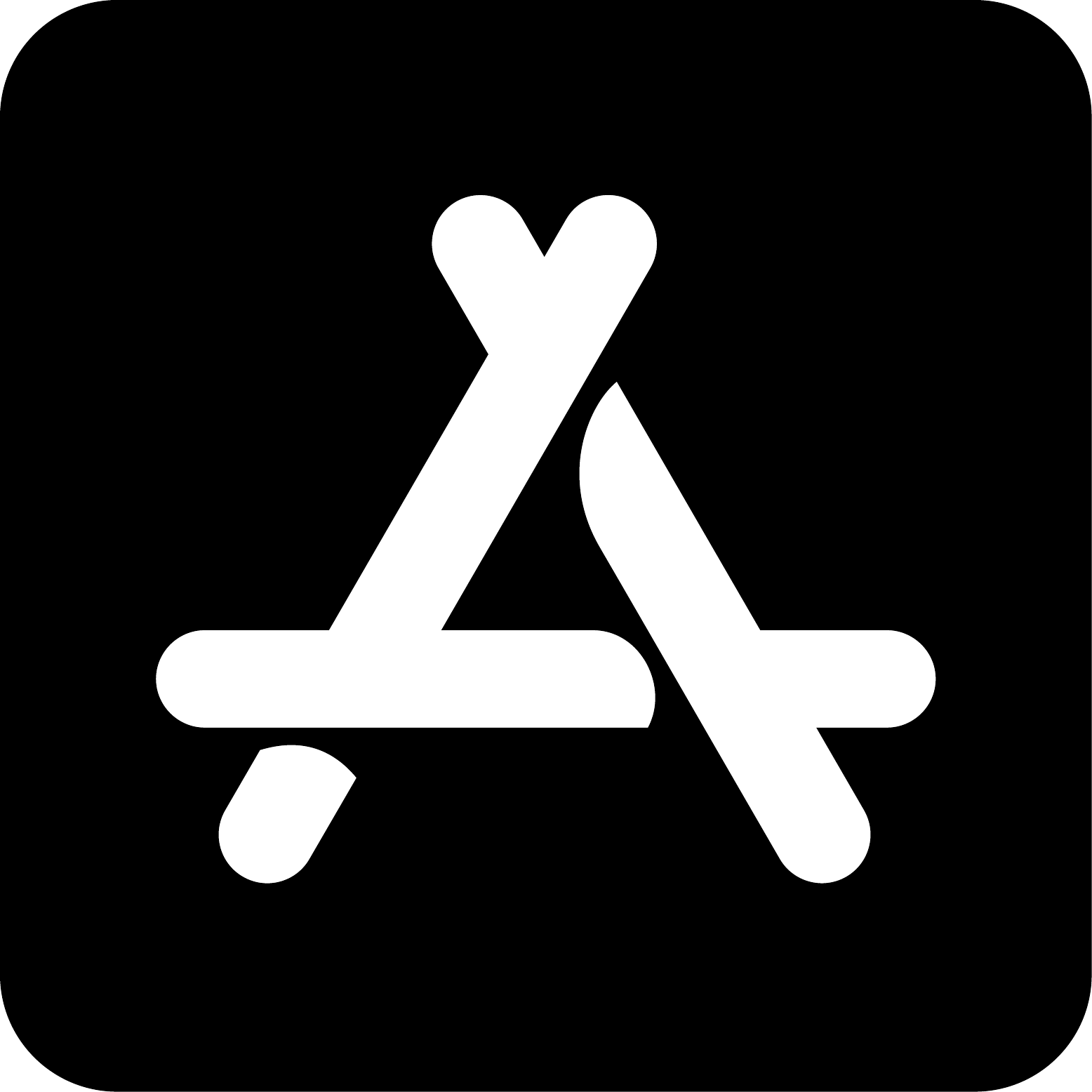}),
	daemons (\includegraphics[height=\fontcharht\font`\B]{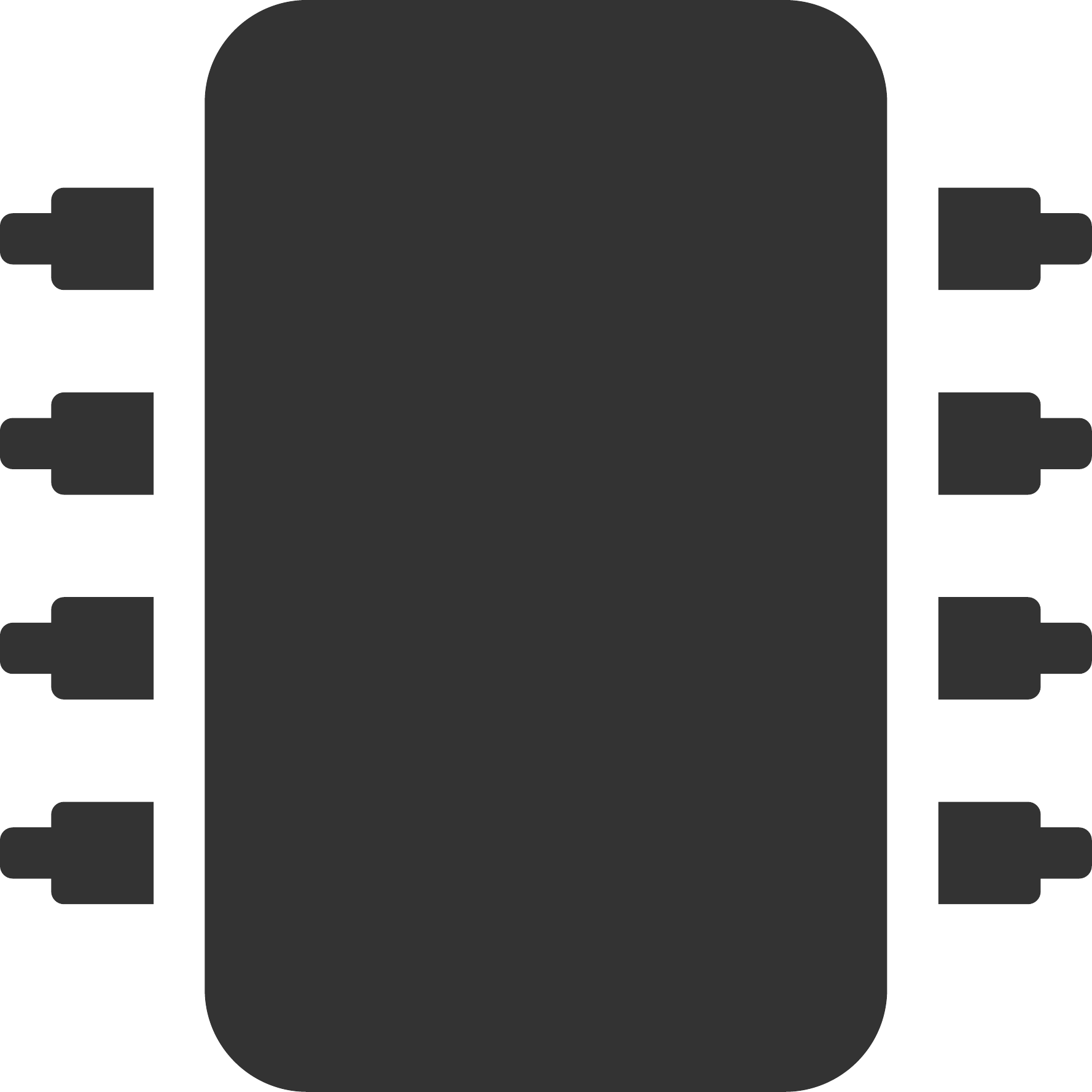}),
	frameworks (\includegraphics[height=\fontcharht\font`\B]{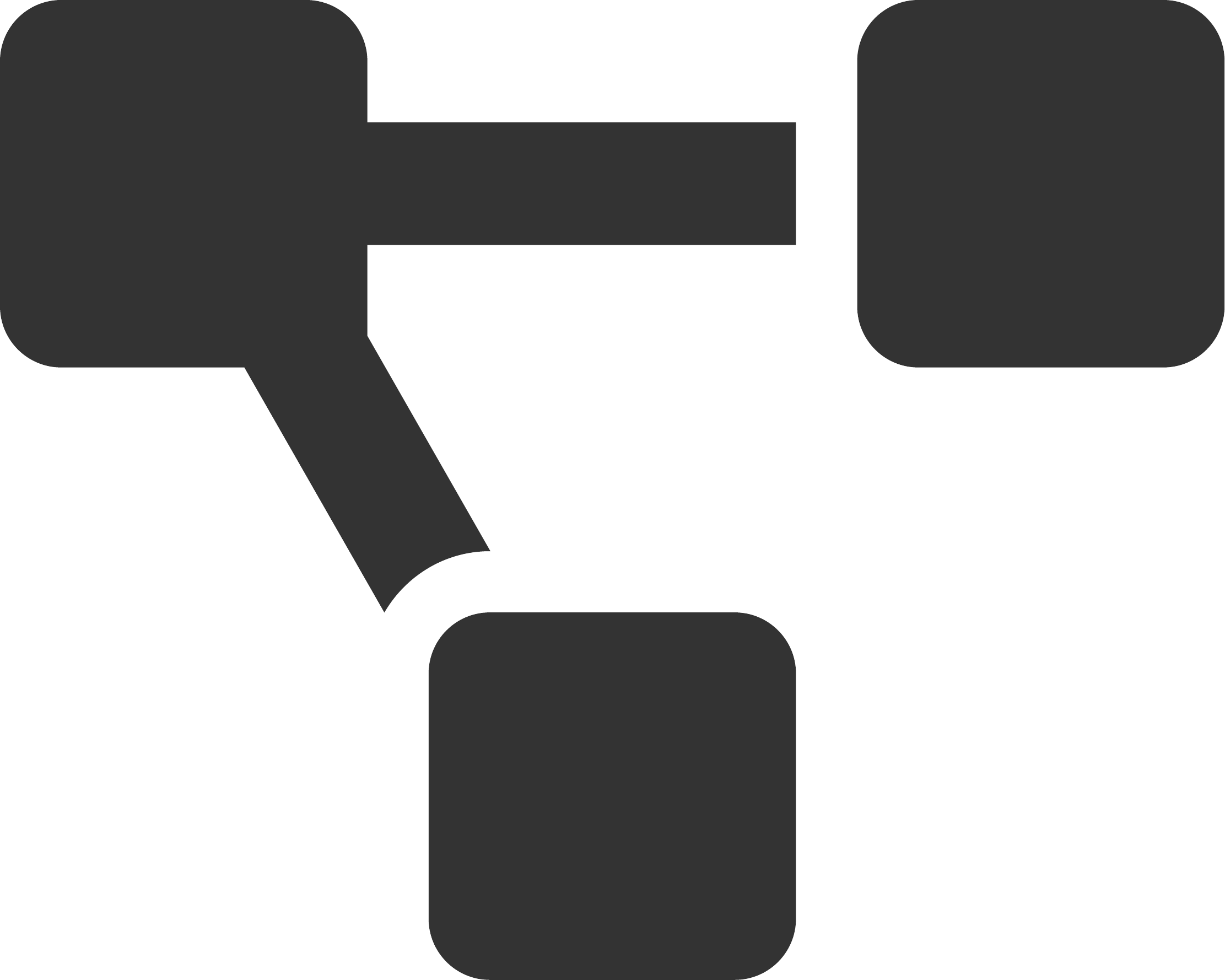}), 
	and drivers (\includegraphics[height=\fontcharht\font`\B]{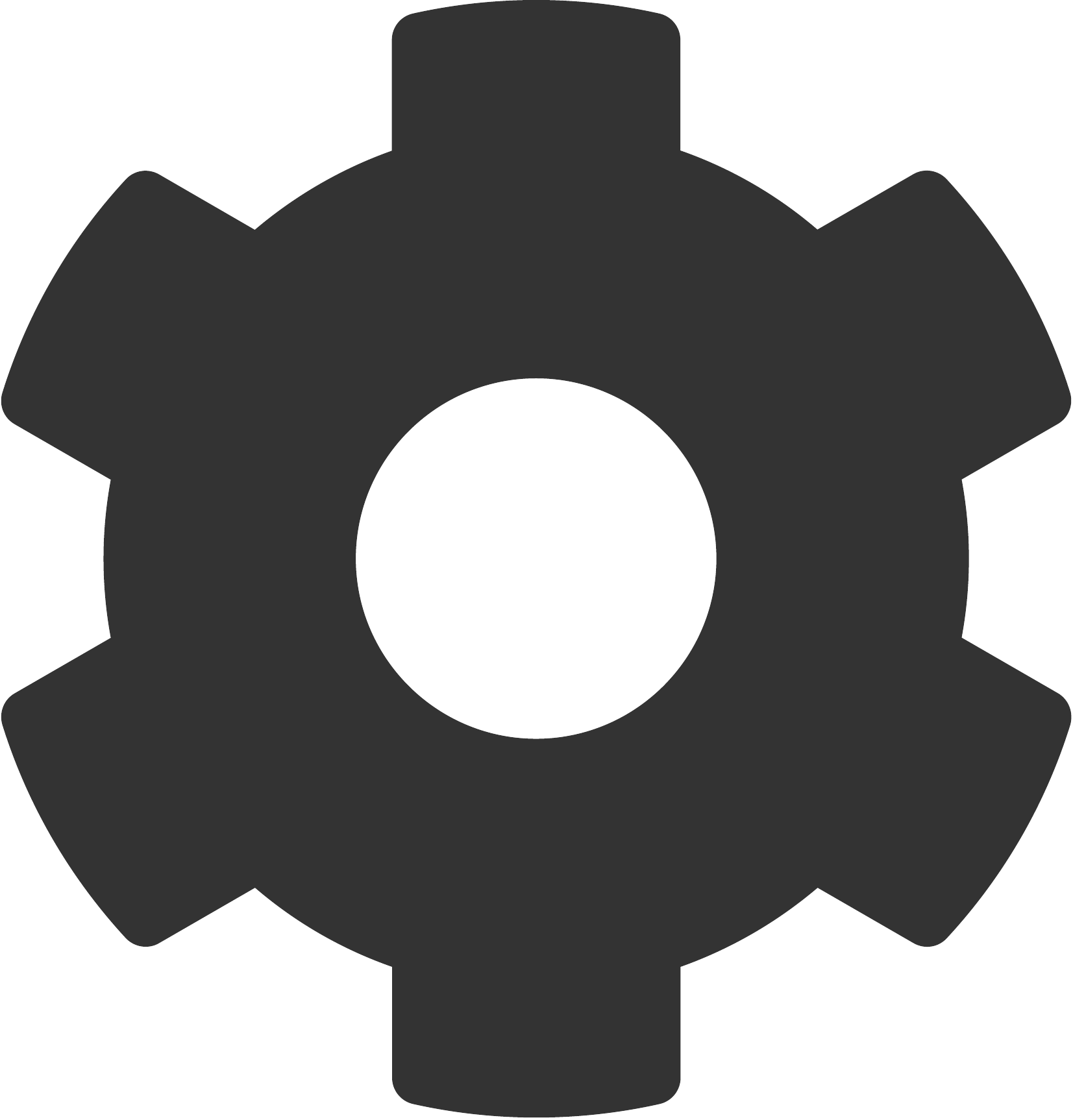}) 
	that are used by \gls{OF} on iOS.
	We highlight the two involved external communication interfaces in \textcolor{mybluetext}{blue}.}
	\label{fig:system_overview}
\end{figure}

Apple's \gls{OF} system is implemented across several daemons and frameworks which communicate via XPC, Apple's implementation of interprocess communication~\cite{Apple:documentation_xpc}. We depict the dependencies of the iOS implementation in \cref{fig:system_overview}. The main daemon that handles \gls{OF} is \emph{searchpartyd}, which runs with root privileges. It generates the necessary keys and performs all cryptographic operations. The daemon is also responsible for communicating with Apple's servers to synchronize keys, submit location reports as a finder device, and fetch location reports as an owner device.
The \emph{bluetoothd} daemon is responsible for sending and receiving \gls{OF} advertisements and passes received advertisements to \emph{locationd}.
The \emph{locationd} daemon adds the device's current location and forwards this information to \emph{searchpartyd}, which generates the finder reports.
On macOS, some functionality of \emph{searchpartyd} such as the server communication is externalized to the \emph{searchpartyuseragent} daemon to support the multi-user architecture that is not available on iOS.

\section{Location Report Accuracy}
\label{sec:evaluation}

We experimentally assess the accuracy of \gls{OF} location reports submitted by finder devices.
This serves two purposes.
\begin{enumerate*}
	\item From an attacker's perspective, we can determine the severity of the discovered vulnerability allowing unauthorized access to location history described in \cref{sec:vuln2_location_access}.
	\item From an end-user's perspective, we can provide empirical evidence on the quality of \gls{OF} location reports when retrieving lost devices.
\end{enumerate*}

Apple's \emph{Find My} application combines multiple location reports to improve accuracy when displaying a location on a map~\cite{Apple:PlatformSecurity_Spring2020}. It does not show the seven-day history of location reports that can be available on Apple's servers.
In this section, we assess the location report accuracy by using this historical location data.
To this end, we use the same PoC implementation presented in \cref{sec:vuln2_location_access} to access the raw location reports for our own devices.
We conduct two sets of experiments in this section.
First, we evaluate the accuracy of \gls{OF} reports for mobility tracking.
Second, we use the seven-day location data history to profile a user's most visited or \enquote{top} locations---which can be abused for user identification.
We open-source all data and evaluation tools forming this section to make our results reproducible~(cf.~\emph{Availability} section).

\paragraph{A Note on Geographic Coordinates}
Throughout this section, we deal with geographic coordinates, their distances, and their projections.
In order to produce meaningful results, we need to use coordinates in the same reference system.
All locations in this paper are latitude and longitude coordinates in the World Geodetic System 1984 (WGS 84)~\cite{EPSG:4326}, which is also used by GPS.
We apply the EPSG:3857 projection~\cite{EPSG:3857} to visualize coordinates on a map (\eg, \cref{fig:of_reports_map}).
When we calculate the distance between two locations, we use the length of a geodesic, \ie, the shortest path between two points on the surface of the ellipsoidal earth~\cite{Karnay2013}.

\subsection{Path Tracking}
\label{sec:eval:tracking}

We compare reported \gls{OF} locations with GPS traces that we record with the tracked device.
We conduct experiments with different means of transportations in and around a metropolitan area.
We measure the error of the raw \gls{OF} reports to the GPS trace and show that we can significantly improve accuracy by applying a scatterplot smoothing algorithm to the raw reports.

\subsubsection{Experimental Setup}

\begin{table}
	\caption{Evaluation scenarios including distance traveled and duration of recorded GPS traces, and the number of downloaded \gls{OF} location reports.}
	\label{tab:location_scenarios}
	\small
	\setlength\tabcolsep{5pt}
	\begin{tabularx}{\linewidth}{Xrrr}
		\toprule
		\textbf{Scenario} &  \textbf{Distance (m)} &        \textbf{Duration (h:m:s)} &  \textbf{No. OF reports} \\
		\midrule
		Walking    &      3375 &  0:55:11 &              489 \\
		Restaurant &       160 &  0:42:29 &              185 \\
		Train      &     23237 &  0:35:30 &              166 \\
		Car        &     94569 &  1:04:22 &               25 \\
		\bottomrule
	\end{tabularx}
\end{table}

We conduct experiments in different scenarios that attempt to emulate common mobility patterns: walking, driving a car, riding a train, and visiting a restaurant. All experiments were conducted in and around the city of Frankfurt am Main, Germany. For each scenario, we record a GPS trace that serves as the ground truth for our evaluation.%
\footnote{The experiments were conducted during the COVID-19 pandemic. Consequently, the authors implemented anti-infection measures, \ie, they wore face masks, used hand sanitizer while traveling with public transport and exercised minimum physical distancing. At the time of the experiments, the local incidence rate was low (five cases per \num{100000} inhabitants over seven days)~\cite{Hessen:CovidInfo}.}

\Cref{tab:location_scenarios} summarizes the evaluated scenarios with the time and distance traveled according to the GPS trace and the number of uploaded \gls{OF} reports during these times.
Our test devices are an iPhone~8 running iOS~13.5 and a MacBook~Pro running macOS~10.15.4 that are logged into the same iCloud account.
During each experiment, we carry the iPhone in flight mode to emit \gls{OF} advertisements and record the GPS trace using the \emph{SensorLog}~\cite{Thomas:SensorLog} application set to a \SI{2}{s} sampling interval.
The MacBook acts as the owner device that we use to download location reports after each experiment.
We did not carry any other Apple devices during an experiment, so we would not get additional reports.
Consequently, all downloaded reports were submitted by the devices of pedestrians.
To access the private advertising keys for downloading and decrypting location reports, we use our PoC implementation (cf.~\cref{sec:vuln2_location_access}).
We use the \gls{OF} reports' generation timestamps to filter the reports that lie between the start and end date of each GPS trace.

\begin{figure}
	\centering
	\includegraphics{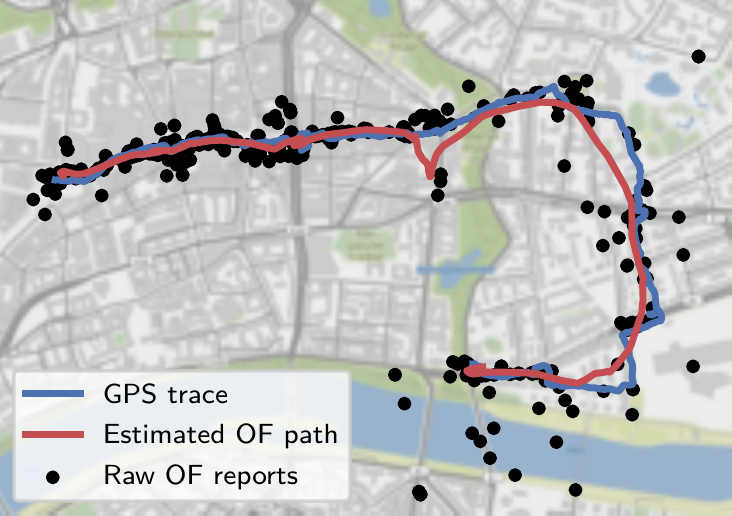}
	\caption{Map of the \emph{walking} scenario showing the GPS trace, the raw OF location reports, and the estimated path calculated from the reports.}
	\label{fig:of_reports_map}
\end{figure}

\subsubsection{Raw Location Report Accuracy}

We calculate the distance of the \gls{OF} reports to the GPS trace.
As GPS trace and \gls{OF} reports do not have a common time index, we interpolate the GPS trace to the time indices of the \gls{OF} reports.
Then, we calculate the distance of each \gls{OF} report to the corresponding point on the interpolated GPS trace.
\Cref{tab:location_error} shows the mean error of the raw reports.
We can see that raw reports have a mean error in the order of \SI{100}{m} for the \emph{walking} and \emph{restaurant} scenarios.
However, results become less accurate when using faster modes of transportations, \eg, \SI{500}{m} when riding a train.
\Cref{tab:location_error} also shows the reported accuracy value, which is part of the \gls{OF} reports (cf.~\cref{sec:protocol_finding}), and the improved results of the estimated path that we will discuss next.

\subsubsection{Estimated Path Accuracy}

\begin{table}
\caption{Accuracy of \gls{OF} location reports.}
\label{tab:location_error}
	\small
	\begin{tabular*}{\linewidth}{l@{}rrrc}
		\toprule
		& \multicolumn{3}{c}{\textbf{Mean distance to GPS trace (m)}} & \textbf{Improvement} \\
		\cmidrule{2-4}
		\textbf{Scenario}   &      \textbf{Reported} &  \textbf{Raw} &  \textbf{Est.\,Path} & \textbf{Raw→Est.\,Path} \\
		\midrule
		Walking    &              121.9 &          81.4 &            25.9 &          $\num{3.1}\times$ \\
		Restaurant &              117.2 &          60.2 &            27.4 &          $\num{2.2}\times$ \\
		Train      &              171.0 &         440.7 &           299.6 &          $\num{1.5}\times$ \\
		Car        &              145.2 &         580.7 &             --- &          --- \\
		\bottomrule
	\end{tabular*}
\end{table}

The raw location reports provide a decent accuracy sufficient to pinpoint an individual's location to a city district or even a street.
However, when plotting the locations on a map (cf.~\cref{fig:of_reports_map}), we see that simply \enquote{connecting the dots} does not yield a smooth path that we would expect from a human mobility trace.
We wondered if we could improve accuracy by harnessing and combining all finders' reports. Essentially, we want to provide a better estimate of the actual path by fitting a curve through the reported \gls{OF} locations.

To this end, we apply a popular curve fitting method called \gls{LOWESS}~\cite{LOWESS} independently to the longitude and latitude coordinates of the location reports.
\gls{LOWESS} performs a weighted local (or moving) regression over a window of nearby data points. A Gaussian distribution assigns a higher weight to the data points in the center of the window.
The size of the window heavily influences the performance of \gls{LOWESS}. Empirically, we found that a value of 30 reports provides the most accurate results for our traces.

\Cref{fig:of_reports_map} shows such an estimated path calculated from the \emph{walking} scenario. The figure shows that our path estimation algorithm approximates the real path rather well.
We quantify the accuracy of the estimated path by calculating the mean distance to the GPS track and show the results in \cref{tab:location_error}.
Most \gls{OF} locations were reported in busy places in a metropolitan city (walking and restaurant visit). 
Here, our fitting algorithm approximated the real path with a mean error below \SI{30}{m}---a $\num{3.1}\times$ improvement over the raw data.
Our fitting algorithm is unable to produce meaningful results for the \emph{car} experiment as the sample set is too sparse.
For completeness, we include maps of the restaurant, train, and car scenario in~\cref{appendix:eval}.

Overall, using a fitting algorithm helps to reconstruct the user's path from noisy \gls{OF} reports \emph{if} the dataset is dense enough.
In the best case, we improved the accuracy compared to the raw location reports by a factor of $\num{3.1}\times$.

\subsection{Identifying Top Locations}
\label{sec:eval:top_locations}

Apple's servers store \gls{OF} reports for seven days, which can be accessed by an unauthorized third-party application (cf.~\cref{sec:vuln2_location_access}).
Previous work has shown that the top (most visited) locations can be used for user fingerprinting.
\textcite{uniqueInTheCrowd} demonstrated that four spatio-temporal points are sufficient to identify \SI{95}{\percent} of all individuals in an anonymized location dataset. Also, \textcite{AnonymizatonDoesNotWork} found that the three top locations on a mobile cell-level are accurate enough to identify \SI{50}{\percent} of all individuals in a large data set.
Even though most devices try to keep a permanent connection to the Internet, our analysis has shown that hundreds of reports have been generated throughout a week.  
This section shows that \gls{OF} reports can also be used to determine top locations and even achieve an accuracy of up to \SI{5}{\m}.

\subsubsection{Experimental Setup}

To determine the top locations, we want to use the same seven-day historic data available as an attacker getting access to the \gls{OF} report decryption keys (cf.~\cref{sec:vuln2_location_access}).
Collecting this data from various test subjects for an evaluation would be extremely intrusive to their privacy.
Running our evaluation algorithm on the subjects' Macs would also not be feasible, because the subjects would have to disable macOS's \gls{SIP} to download raw \gls{OF} location reports, effectively weakening their system security.
Sensibly, our ethical review board (ERB) would not approve such a study.
Consequently, we abstain from conducting a user study and, instead, use our (the authors') data for the evaluation.
To preserve the authors' privacy, we will not show or discuss the raw data.
Instead, we apply our algorithm for identifying top locations on the raw data and then discuss the output.

\subsubsection{Resampling and Clustering OF Reports}

Previous works~\cite{uniqueInTheCrowd,AnonymizatonDoesNotWork} have used identifiers of cell sites to rank top locations. For a particular user, they simply count the number of registrations per cell and base the location rank on this number. \Eg, the cell with the highest registration count is regarded as the top 1 location. We cannot apply the same concept for identifying top locations based on \gls{OF} reports for two reasons.

\paragraph{Problem 1: Non-Uniform Distribution Over Time}
We observed that the density of \gls{OF} reports over time is non-uniform.
In busy places such as malls or restaurants, more finder devices are available and, consequently, more reports are generated.
If we simply count the number of \gls{OF} reports to determine the top locations, these busy places are likely to be overrepresented.
Instead, we want to rank top locations based on the overall visit duration of the user.

\paragraph{Problem 2: Continuous Coordinate Range}
Cell identifiers are elements of a discrete set. In contrast, the location coordinates in \gls{OF} reports are drawn from a continuous range, which makes simple counting of visits per location hard.
To address both issues, we use a two-step approach.

\paragraph{Solution 1: Resampling}
We resample the location reports on the time axis to solve Problem 1 and \enquote{flatten} the distribution over time.
Within each resampling interval $R$, we calculate the center coordinates as the mean of all reports within that interval.
Setting a reasonable value $R$ is essential.
If $R$ is too small, the desired flattening effect will not occur.
If $R$ is too large, we lose accuracy.
Empirically, we found that $R = \SI{20}{min}$ produces good results for our dataset.

\paragraph{Solution 2: Clustering}
We use a clustering algorithm to identify places for which we received multiple location reports over time to solve Problem 2.
In particular, we select the popular Density-Based Spatial Clustering of Applications with Noise (DBSCAN)~\cite{Ester1996:DBSCAN,Schubert2017:DBSCAN} algorithm.
DBSCAN can detect an arbitrary number of clusters, which is important for us as we do not know the number of top locations a priori. Also, DBSCAN adequately deals with noise, which is common in \gls{OF} reports.
In short, DBSCAN forms clusters by finding \enquote{core samples} that have at least $N$ neighbors within a radius of $D$. All samples that are not part of a cluster are considered as noise.
Empirically, we determined that $D=\SI{50}{\m}$ and $N=6$ produces good results for our dataset.

\begin{table}
\caption{Identified top locations sorted by their rank. Error indicates the geodesic distance between the ground truth and the estimated cluster center. Days indicate on how many days of the week the location was visted. Dwell time is the estimated overall time that the location was visited.}
\label{tab:top_locations}
	\small
	\begin{tabularx}{\linewidth}{lcrYrY}

	\toprule
	\textbf{Kind} & \textbf{Rank} & \textbf{Error} & \textbf{Resampled reports} & \textbf{Days} & \textbf{Dwell time (hour:min)} \\
	\midrule
    Home &     1 &  \SI{14.1}{m} &      129 &             6 & 43:00 \\
    Work &     2 &   \SI{4.9}{m} &       25 &             2 & 08:20 \\
 Partner &     3 &  \SI{15.5}{m} &       19 &             2 & 06:20 \\
 Friends &     4 &  \SI{11.1}{m} &       15 &             1 & 05:00 \\
   Sport &     5 &   \SI{9.5}{m} &        9 &             3 & 03:00 \\
  Family &     6 &   \SI{6.6}{m} &        9 &             2 & 03:00 \\
	\bottomrule

	\end{tabularx}
\end{table}

\subsubsection{Results}

We run our resampling and clustering algorithms on an author dataset.
Together, the algorithms determine a list of six clusters that we interpret as top locations.
Afterwards, we let the author label each entry and assign the actual location of the home or office by picking coordinates from an online map service, which we use as ground truth.
Finally, we compute the distance of the clusters' centers to the ground truth locations and show the complete results in~\cref{tab:top_locations}.
The results demonstrate that we successfully identified typical locations such as home and workplace~\cite{AnonymizatonDoesNotWork} with an error below~\SI{20}{\m}.
Impressively, the location reports were precise enough to pinpoint not only the workplace but with a \SI{5}{\m} error even identify the exact office.

\begin{figure}
	\centering
	\includegraphics{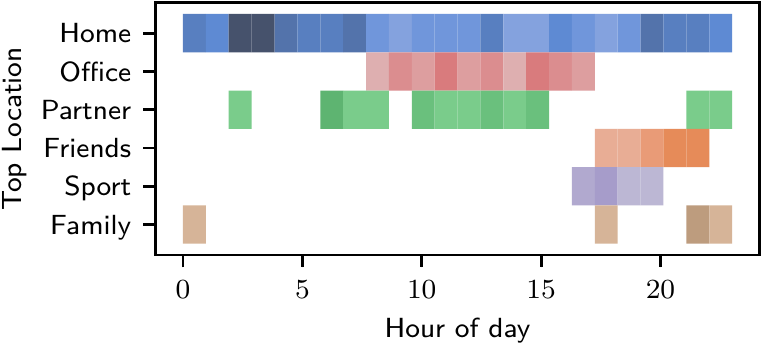}
	\caption{Visiting times distribution of top locations averaged over seven days. Darker colors indicate a higher number of location reports during these hours.}
	\label{fig:toplocations}
\end{figure}

We wondered if it would have been possible to discern the type of location (\eg, home or work) just from the characteristics of the location reports.
To this end, we provide a detailed view of the visiting times across the hours of a day in~\cref{fig:toplocations}.
The figure shows the hours of a day in which location reports were submitted for the respective top location.
We can visually identify the type of some locations from this distribution.
In particular, the workplace distribution matches typical office hours (8\,am to 5\,pm).
The all-day distribution of the home location reflects the fact that the measurements were taken over the course of a week and the author stayed at home on some days for remote work. 

Previous work showed that four spatio-temporal points could be sufficient to uniquely identify an individual with a chance of \SI{95}{\percent}~\cite{uniqueInTheCrowd}.
While we were unable to conduct a similar large-scale user study as in~\cite{uniqueInTheCrowd} based on \gls{OF}, our small-scale experiment already demonstrates that \gls{OF} reports contain highly sensitive information that can be used to deanonymize users.

\begin{table*}
	\newcommand{\vhigh}{++}
	\newcommand{\high}{+}
	\newcommand{\med}{o}
	\newcommand{\low}{-}
	\newcommand{\vlow}{{-}-}
	\caption{Summary of our security and privacy analysis of \gls{OF}. We list potential issues and provide a short assessment about whether or not they are exploitable in practice (\cmark or \xmark) and if \cmark, by which adversary model (cf.~\cref{sec:adversary_model}).}
	\label{tab:analysis}
	\small
	\begin{tabularx}{\linewidth}{lp{2.7cm}cX}
		\toprule
		\textbf{Component} & \textbf{Potential issue}  & \textbf{Exploit.} & \textbf{Assessment} \\
		\midrule

		Cryptography
			& Key diversification  & \xmark & The custom key diversification process follows the NIST recommendation for key derivation through extraction-then-expansion~\cite{NIST:KDF}. \\
			& Choice of P-224 curve  & \xmark & Use of NIST P-224 is discouraged by some cryptographers~\cite{SafeCurves}. However, we are unaware of any practical attacks against P-224 when used exclusively for \gls{ECDH}. \\
			& Insecure key storage  & \cmark~\textbf{(A1)} & Keychains and SEP are used to securely store keys for server communication and the master beacon key. However, macOS caches the derived advertisement keys on disk, which can be read by local applications. Attackers can exploit this to access (historical) geolocation data as we describe in \cref{sec:vuln2_location_access}. \\

		\midrule

		Bluetooth    
			& Device tracking via \gls{BLE} advertisements  & \xmark & BLE payload and address are determined by the advertisement key, which is changed at \SI{15}{min} intervals, making long-term tracking hard. \\
			& Remote code execution (RCE)  & \xmark & As \gls{OF} uses non-connectable mode to emit advertisements, devices remain secure against RCE attacks on the Bluetooth firmware~\cite{Ruge2020}. \\
			& Denial-of-service (DoS)  & \cmark~\textbf{(A2)} & An attacker could emit/relay legitimate advertisements at other physical locations to pollute the set of location reports. \\

		\midrule

		Server comm.
			& Spoofing (finder)  & \xmark & Impact similar to Bluetooth relaying. However, we have been unable to inject fabricated location reports into the server communication. \\
			& Spoofing (owner)  & \xmark & Spoofing an owner device is not critical as location reports are end-to-end encrypted. \\
			& Device identification  & \cmark~\textbf{(A4)} & Apple's servers can identify both finder and owner devices. This enables a location correlation attack that we discuss in \cref{sec:vuln1_correlate}. \\
		\bottomrule
	\end{tabularx}
\end{table*}

\section {Security and Privacy Analysis}
\label{sec:security_analysis}

In this section, we perform a security and privacy analysis of Apple's \gls{OF} system implemented on iOS and macOS based on the adversary models described in \cref{sec:adversary_model}.
We first examine the cryptography-related components that are relevant for the local application~\textbf{(A1)} and service operator~\textbf{(A4)} models that have access to keys and encrypted reports, respectively. Then, we assess the BLE interface relevant to the proximity-based adversary~\textbf{(A2)} and the HTTPS-based server communication relevant for the network-based adversary~\textbf{(A3)}. We summarize our findings in \cref{tab:analysis} and discuss in the following.%

\subsection{Cryptography}

\paragraph{Key Diversification}
\gls{OF} employs key diversification to derive the rolling advertisement keys from the master beacon key~(cf.~\cref{sec:Cryptography}). Apple's design follows the NIST recommendation of performing extraction-then-expansion~\cite{NIST:KDF} to securely derive keys.
The two-step process first extracts a derivation key from a secure input and then expands this key to the desired output length. Specifically, \gls{OF} first extracts a new 32-byte key $SK_i$ from the previous derivation key using the KDF and then expands $SK_i$ using the same KDF to 72 bytes.

\paragraph{Choice of NIST P-224 Curve}
We believe that Apple's choice for the NIST P-224 curve is the consequence of the constrained capacity of \gls{BLE} advertisements while maximizing the security level of the encryption keys. 
Apple's implementation of P-224 in \emph{corecrypto} has been submitted to validate compliance with \gls{FIPS}~\cite{Apple:Security}.
Within the cryptography community, some researchers discourage the use of P-224 because its generation process is unclear~\cite{Bernstein2006,SafeCurves}. 
More modern curves with the same security margin are available, \eg, M-221~\cite{cryptoeprint:2013:647}, but are not used by Apple.

\paragraph{Insecure Key Storage}
We analyzed how \gls{OF} keys and secrets are stored on the system. While most involved keys are synchronized and stored in the iCloud keychain, we discovered that the advertisement keys derived from the master beacon key~(cf.~\cref{sec:Cryptography}) are cached on disk to avoid unnecessary re-computations. We found that the cached key directory is accessible by a local application with user privileges and can be used to bypass the system's location API, as we describe in \cref{sec:vuln2_location_access}.

\subsection{Bluetooth}

\paragraph{Device Tracking}
One of the key design goals of \gls{OF} is to prevent tracking of lost devices via their \gls{BLE} advertisements.
According to our analysis, \gls{OF} fulfills this promise by randomizing both \gls{BLE} advertisement address and payload in \SI{15}{min} intervals~(cf.~\cref{sec:protocol_losing}).

\paragraph{Remote Code Execution}
In addition, \gls{OF} uses the so-called \enquote{non-connectable mode}~\cite[Vol.~3, Part~C, §~9.3.2]{Bluetooth5_2:CoreSpec}, which means that other devices cannot connect to it and exploit potential remote code execution (RCE) vulnerabilities in the Bluetooth firmware~\cite{Ruge2020}.

\paragraph{Denial-of-Service Through Relaying}
\gls{BLE} advertisements only contain the public part of an advertisement key and are not authenticated.
Anyone recording an advertisement can replay it at a different physical location. Any finder at that location would generate a location report and submit it to Apple.
Through this type of relaying, an attacker could make a lost device appear at a different location, effectively mounting a \gls{DoS} attack as owners would receive different contradicting location reports.

\subsection{Server Communication}

\paragraph{Spoofing}
The communication with Apple's servers uses \gls{TLS}, including certificate pinning to ensure that no \gls{MitM} attack can be deployed.
Based on our analysis, the protocol seems to implement a secure authentication scheme. However, we have been unable to reconstruct some of the involved components.
We understand that a device-specific certificate (cf.~\cref{sec:protocol_finding}) and a private signing key, protected by the \gls{SEP}, are involved in submitting reports.
We \emph{assume} that this private key is used for remote attestation to prevent non-Apple devices from submitting potentially fabricated reports. The generation and registration process of these keys with Apple's server remains unknown to us. 
Also, the \enquote{anisette data} used for authenticating owner devices (cf.~\cref{sec:protocol_searching}) is not publicly documented, and the code that generates the tokens is highly obfuscated.

\paragraph{Device Identification}
While we did not recover the exact details of the authentication mechanism, we have observed that both finder and owner devices provide identifiable tokens to Apple's servers. In particular, owner devices provide their Apple ID to access location reports.
In \cref{sec:vuln1_correlate}, we show that by requesting IDs, Apple's servers are---in principle---able to correlate the locations of different owners.


\section{Apple Can Correlate User Locations}
\label{sec:vuln1_correlate}

Apple as the service provider~\textbf{(A4)} could infer that two or more owners have been in close proximity to each other as \gls{OF} uses identifiable information in both upload and download requests.
Law enforcement agencies could exploit this issue to deanonymize participants of (political) demonstrations even when participants put their phones in flight mode.
Exploiting this design vulnerability requires that the victims request the location of their devices via the Find My application.\footnote{This requirement currently limits the exploitability of the described vulnerability. However, exploitability could increase when changing the usage pattern of the downloading API~(cf.~\cref{sec:protocol_searching}). For example, a future Apple application could notify users about lost devices automatically by regularly requesting reports for the devices' last known locations, thereby, removing the need for user interaction.}
Next, we describe the vulnerability, a possible attack, and our proposed mitigation.

\subsection{Vulnerability}

When uploading and downloading location reports, finder and owner devices reveal their identity to Apple.
During the upload process, the finder reveals a device-specific identifier in the HTTPS request header (cf.~\cref{tab:submit_reports_headers}) that can be used to link multiple reports to the same finder.
Similarly, during the download process, the owner device has to reveal its Apple ID. In particular, the owner includes its Apple ID in the HTTPS request headers (cf.~\cref{tab:fetch_reports_headers}), which allows Apple to link reports uploaded by a particular finder to the Apple ID of the downloading owners.
Since we do not have access to Apple's servers, we cannot make assumptions about whether or not Apple actually stores such metadata. However, the fact that Apple \emph{could} store this information indefinitely opens the possibility of abuse.

\subsection{Attack}

\begin{figure}
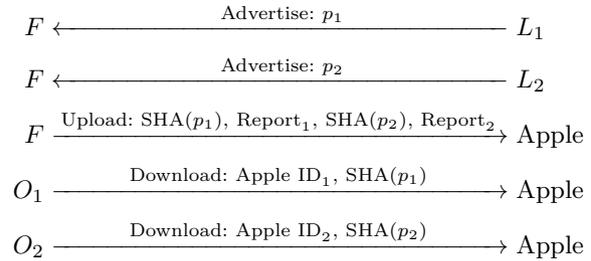

	\begin{align}
		F & \xleftarrow{\mathmakebox[5.8cm]{\text{Advertise: } p_1}} L_1 \nonumber \\
		F & \xleftarrow{\mathmakebox[5.8cm]{\text{Advertise: } p_2}} L_2 \nonumber \\
		F & \xrightarrow{\mathmakebox[5.8cm]{\text{Upload:} \text{ SHA}(p_1), \text{ Report}_1, \text{ SHA}(p_2), \text{ Report}_2}} \text{Apple} \nonumber \\
		O_1 & \xrightarrow{\mathmakebox[5.8cm]{\text{Download:} \text{ Apple ID}_1, \text{ SHA}(p_1)}} \text{Apple} \nonumber \\
		O_2 & \xrightarrow{\mathmakebox[5.8cm]{\text{Download:} \text{ Apple ID}_2, \text{ SHA}(p_2)}} \text{Apple} \nonumber
	\end{align}
	\caption{Apple could infer which users have been in close proximity to each other.}
	\label{fig:correlation}
\end{figure}

It is possible for Apple to find out which owners have been in physical proximity to each other \emph{if the owners request the location of their devices via the Find My application.}
We sketch the attack for two owners in \cref{fig:correlation}.
A finder $F$ receives advertisements from the lost devices $L_1$ and $L_2$ that belong to the owners $O_1$ and $O_2$, respectively, and publishes encrypted location reports to Apple's servers.
Due to the limited communication range of \gls{BLE}, we can reasonably assume that $L_1$ and $L_2$ have been in close proximity if the respective location reports were generated at a similar time and submitted by the same finder.
Later, $O_1$ and $O_2$ both download location reports, by opening the \emph{Find My} app, for $L_1$ and $L_2$, respectively.
At this point, Apple can infer that these two owners identified by their Apple IDs were close to each other.

\subsection{Impact}

The presented attack could be harmful to protesters who put their phones into flight mode to stay anonymous and prevent their devices from showing up during a cell site analysis---which is precisely when the devices would start emitting \gls{OF} advertisements. Law enforcement agencies could record all the advertised public keys at the demonstration site and ask Apple to provide the Apple IDs of the users that later requested location reports to deanonymize the participants.
Such a collusion would be a combination of the proximity-based~\textbf{(A2)} and service provider~\textbf{(A4)} adversary models~(cf.~\cref{sec:adversary_model}).

\subsection{Proposed Mitigation}

There are two straightforward options to mitigate this attack: remove identifying information from either (1) finder devices or (2) owner devices.
We assume that the authentication of the finder provides a form a remote attestation proving that the device is---in fact---a genuine Apple device allowed to upload location reports to Apple's servers. In that case, option (1) is not feasible as the finder has to provide some verifiable information by design.
However, we currently see no reason why owner devices have to authenticate to Apple's servers and provide personally identifiable information, \ie, the Apple ID.
We found that any Apple device can request arbitrary location reports, so the authentication appears to be a security-by-obscurity measure and only prevents everyone without access to an Apple device from accessing location reports.
Therefore, we recommend option (2) as mitigation and disable authentication for download requests.


\section{Unauthorized Access of Location History}
\label{sec:vuln2_location_access}

We discovered a vulnerability of the \gls{OF} implementation on macOS that allows a malicious application~\textbf{(A1)} to effectively circumvent Apple's restricted location API~\cite{Apple:CoreLocation} and access the geolocation of all owner devices without user consent. Moreover, historical location reports can be abused to generate a unique mobility profile and identify the user, as we demonstrate in \cref{sec:evaluation}.

\subsection{Vulnerability}

\Cref{sec:protocol} describes that the location privacy of lost devices is based on the assumption that the private part of the advertisement keys is only known to the owner devices.
The advertisement keys change every 15 minutes and \gls{OF} supports retrieving location reports from the last seven days, so there is a total of 672 advertisement keys per device, for which there exist potential location reports on Apple's servers.
In principle, all of these keys could be generated from the master beacon key~(cf.~\cref{sec:Cryptography}) whenever needed. However, Apple decided to cache the advertisement keys, most likely for performance reasons.
During our reverse engineering efforts, we found that macOS stores these cached keys on disk in the directory \path{/private/var/folders/<Random>/com.apple.icloud.searchpartyd/Keys/<DeviceId>/Primary/<IdRange>.keys}.
The directory is readable by the local user and---in extension---by any application that runs with user privileges.
On iOS, those cache files exist as well, but they are inaccessible for third-party applications due to iOS's sandboxing mechanism.

\subsection{Attack}

\begin{figure}
\includegraphics[width=\linewidth]{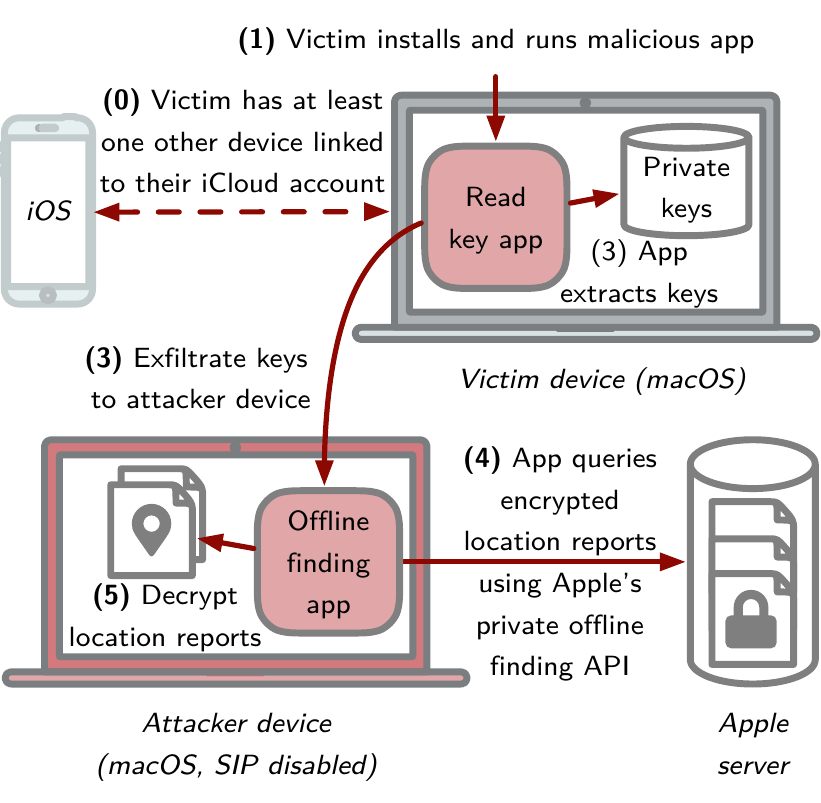}
\caption{Attack flow for gaining access to the victim's location history. Attacker-controlled components are marked in \textcolor{myredtext}{red}.}
\label{fig:poc}
\end{figure}

We describe the attack flow and explain our PoC implementation, which leads to the attacker gaining access to the location history of the victim's devices. In the following, we detail the operation of our two-part PoC attack. The steps are referring to \cref{fig:poc}. 

\paragraph{Reading Private Keys (Steps 1--3)}
The victim installs a non-sandboxed malicious application.\footnote{Sandboxing is only required for applications distributed via Apple's app store. Many popular macOS applications such as Firefox or Zoom are not distributed via the app store and, thus, could have exploited the discovered vulnerability.} When started, the malicious application runs with user privileges and, therefore, has access to the key cache directory. It can read the advertisement keys from disk (2) and then exfiltrate them to the attacker's server (3).
Apart from starting the application, this process requires no user interaction, \ie, no dialogs requesting disk access are displayed to the user.

\paragraph{Downloading Location Reports (Step 4)}
The \emph{attacker's machine} essentially acts as an owner device (cf.~\cref{sec:protocol_searching}) but uses the victim's keys as input for the HTTPS download request.
To download the victim's location reports, our PoC needs to access the attacker's \emph{anisette data} for authenticating the request to Apple's servers.  
As we need to link private frameworks and access the anisette data in our implementation, the attacker's macOS system needs to disable \gls{SIP} and \gls{AMFI}.
Since this device is attacker-owned, this requirement does not limit the applicability of the presented attack.
\gls{SIP} and \gls{AMFI} are disabled by booting in the macOS recovery mode and running the following terminal commands. 
\begin{lstlisting}
csrutil disable
nvram boot-args="amfi_get_out_of_my_way=1"
\end{lstlisting}

\paragraph{Decrypting Location Reports (Step 5)}
In the final step, the adversary uses the victim's private keys to decrypt the location reports.

\subsection{Impact}

The attack essentially allows any third-party application to \emph{bypass Apple's Core Location API}~\cite{Apple:CoreLocation} that enforces user consent before an application can access the device's location.
Moreover, the attacker can access the location history of the past seven days of \emph{all} the owner's devices.
The victim is only required to download and run the application but remains otherwise clueless about the breach.
Our analysis has shown that the advertisement keys are precomputed for up to \emph{nine} weeks into the future, which allows an adversary to continue downloading new reports even after the victim has uninstalled the malicious application. 

Even though the location reports are not continuous, our evaluation in \cref{sec:evaluation} shows that it is easy to identify the user's most visited places such as home and workplace. Furthermore, we show that the decrypted location reports can accurately track the victim's movement of the last seven days.

\subsection{Mitigation}

As a short-term mitigation, users can disable participating in the \gls{OF} network to prevent the attack.
In addition, we propose three long-term solutions to mitigate the attack:
\begin{enumerate*}
	\item encrypting all cached files on disk store the decryption key in the keychain,
	\item restricting access to the cache directory via access control lists,
	\item not caching the keys and computing them on-demand. 
\end{enumerate*}
In fact, macOS~10.15.7 includes a mitigation based on option (2), which moved the keys to a new directory that is protected via the system's sandboxing mechanism.%


\section{Related Work}
\label{sec:relatedwork}

We review other crowd-sourced location tracking systems and previous security and privacy analyses of Apple's ecosystem.

\paragraph{Crowd-Sourced Location Tracking}
\textcite{Weller_2020} have studied the security and privacy of commercial Bluetooth tags (similar to Apple's definition of \emph{accessories}) sold by multiple vendors.
Many of the studied systems provide crowd-sourced location tracking similar to Apple's \gls{OF}, allowing users to discover lost devices by leveraging  the finder capabilities of other devices.
The study discovered several design and implementation issues, including but not limited to the use of plaintext location reports, unauthorized access to location reports, broken \gls{TLS} implementations, and leaking user data.
Based on their findings, \textcite{Weller_2020} propose a novel privacy-preserving crowd-sourced location tracking system called \emph{PrivateFind}. PrivateFind does not need user accounts and uses end-to-end encrypted location reports with a symmetric encryption key stored on the Bluetooth finder during the initial setup. In their solution, a finder that discovers a lost Bluetooth tag sends its geolocation to the finder over Bluetooth. The lost device encrypts the location with its symmetric key and returns the encrypted report. The finder then uploads the encrypted location report on behalf of the tag. An owner device that knows the symmetric key can then download and decrypt the location report.

To the best of our knowledge, PrivateFind is the only other privacy-friendly offline device finding system next to \gls{OF}. Both designs achieve similar privacy goals, such as preventing a third party from learning the location.
The main difference is the way encrypted location reports are generated.
\gls{OF} employs public-key cryptography, which allows finder devices to generate a location report upon receiving a single Bluetooth advertisement.
In PrivateFind, lost devices are actively involved in the generation, which leads to the following practical issues:
\begin{enumerate*}
	\item Lost devices or tags drain their batteries quicker as they have to establish Bluetooth connections with other devices and perform cryptographic operations. This opens up the door for resource-exhaustion attacks where a powerful adversary issues an excessive number of encryption requests to the lost devices.
	\item The lack of finder attestation would allow an attacker to upload fabricated reports as the lost device cannot verify the correctness of the provided location.
\end{enumerate*}

\paragraph{Apple's Wireless Ecosystem Security and Privacy}
Previous work analyzed parts of Apple's wireless services.
\textcite{Bai2016} investigated the risks of using insecure \gls{mDNS} service advertisements and showed that they have been able to spoof an AirDrop receiver identity to get unauthorized access to personal files.
\textcite{Stute2018,Stute2019} reverse engineered the complete \gls{AWDL} and AirDrop protocols and demonstrated several attacks, including user tracking via \gls{AWDL}, a \gls{DoS} attack on \gls{AWDL}, and a \gls{MitM} attack on AirDrop.
\textcite{Martin2019} extensively analyzed the content of the \gls{BLE} advertisements for several Apple services. They found several privacy-compromising issues, including device fingerprinting and long-term device and activity tracking.
\textcite{Celosia2020} extended this work and discovered new ways of tracking \gls{BLE} devices such as Apple AirPods and demonstrated how to recover user email addresses and phone numbers from \gls{BLE} advertisements sent by Apple's \gls{PWS}.
\textcite{Heinrich2021} found that AirDrop also leaks user phone numbers and email addresses and proposes a new protocol based on private set intersection.
\textcite{Stute2021} investigated the protocols involved in \gls{PWS} and Apple's Handoff and found vulnerabilities allowing device tracking via Handoff advertisements, a \gls{MitM} attack on \gls{PWS}, and \gls{DoS} attacks on both services.
While the above works have analyzed other services, we leveraged their methodology for approaching our analysis and reverse engineering work of \gls{OF}. 


\section{Conclusion}
\label{sec:conclusion}

Apple has turned its mobile ecosystem into a massive crowd-sourced location tracking system called \gls{OF}.
In this system, all iPhones act as so-called finder devices that report the location of lost devices to their respective owners.
Apple claims to implement \gls{OF} in a privacy-preserving manner.
In particular, location reports are inaccessible to Apple, finder identities are concealed, and \gls{BLE} advertisements cannot be used to track the owner~\cite{Apple:BlackHat2019}.
We have been the first to challenge these claims and provide a comprehensive security and privacy analysis of \gls{OF}.

The good news is that we were unable to falsify Apple's specific claims.
However, we have found that \gls{OF} provides a critical attack surface that seems to have been outside of Apple's threat model.
Firstly, the \gls{OF} implementation on macOS allows a malicious application to effectively bypass Apple's location API and retrieve the user's location without their consent. By leveraging the historical reports, an attacker is able to identify the user's most visited location with sub-\SI{20}{m} accuracy.
Secondly, we believe that Apple has yet to provide a good reason why owner devices need to authenticate when retrieving encrypted location reports as it allows Apple to correlate the locations of different Apple users.

We were only able to publish our findings by intensively studying the \gls{OF} system using reverse engineering, which is a very time-consuming process (we started analyzing \gls{OF} mid-2019).
To protect user privacy, we believe that systems handling highly sensitive information such as \gls{OF} need to be \emph{openly and fully} specified to facilitate \emph{timely} independent analyses.
To this end, we urge manufacturers to provide not only partial~\cite{AppleFindMyNetworkSpecification} but complete documentation of their systems and release components as open-source software whenever possible, which is already a best practice for cryptographic libraries~\cite{Apple:Security}.


\section*{Responsible Disclosure}

We disclosed the vulnerability in \cref{sec:vuln2_location_access} on \DTMdate{2020-07-02}. On \DTMdate{2020-10-05}, Apple informed us that macOS~10.15.7 provides a mitigation for the issue, which was assigned CVE-2020-9986.
In addition, we informed Apple about the vulnerability in \cref{sec:vuln1_correlate} on \DTMdate{2020-10-16}, and are currently waiting for feedback.


\section*{Availability}
\label{sec:availability}

\newcommand{\shorturl}[2]{\href{#1}{\nolinkurl{#2}}}

We release the following open-source software artifacts as part of the Open Wireless Link project~\cite{owlink-project}:
\begin{enumerate*}
	\item The PoC implementation that can download and decrypt location reports, which we used for the exploit described in \cref{sec:vuln2_location_access} (\shorturl{https://github.com/seemoo-lab/openhaystack}{github.com/seemoo-lab/openhaystack}).
	\item The experimental raw data and evaluation scripts to reproduce the results in \cref{sec:evaluation} (\shorturl{https://github.com/seemoo-lab/offline-finding-evaluation}{github.com/seemoo-lab/offline-finding-evaluation}).
\end{enumerate*}


\section*{Acknowledgments}

We thank our anonymous reviewers and our shepherd Santiago Torres-Arias for their invaluable feedback.
We thank \shorturl{https://fontawesome.com}{Fontawesome} for the vector graphics and \shorturl{https://stamen.com}{Stamen} for the map tiles used in our figures.
This work has been funded by the LOEWE initiative (Hesse, Germany) within the emergenCITY center and by the German Federal Ministry of Education and Research and the Hessen State Ministry for Higher Education, Research and the Arts within their joint support of the National Research Center for Applied Cybersecurity ATHENE.

\printbibliography


\appendix

\section{HTTP Body Content}
\label{appendix:http}

\Cref{lst:fetch,lst:fetch_response} show the HTTP request and response body for downloading location reports, respectively.

\begin{lstlisting}[caption={Request body for downloading location reports.},label={lst:fetch}, linewidth=\linewidth]
{
  "search": [
    {
      "endDate": 1599052814928,
      "startDate": 1598965514928,
      "ids": [
        "tEJGn1j59g+mgj7cKhDMYN3UMNb8...",
        "sr74jRoVkhXdshf0Y68j6qGyW68v...",
        "pzcyP8dXfdSyVTHk8io7AUgAx85J...",
        ... ]
    }, ... ]
}
\end{lstlisting}

\begin{lstlisting}[caption={Response body for downloading location reports.},label={lst:fetch_response}, linewidth=\linewidth]
{
  "results": [
    {
      "datePublished": 1586804587284,
      "payload": "JETtmwIEzRBG....",
      "description": "found",
      "id": "B6E5tpUPbuudAc..."
      "statusCode": 0
    }, ... ],
  "statusCode": "200"
}
\end{lstlisting}

\section{Reporting Delay}
\label{sec:reporting-delay}

\Cref{fig:reporting-delay} shows the distribution of the reporting delays (time between uploading and generating a report) over all traces recorded for the experiments in~\cref{sec:eval:tracking}.

\section{Additional Experimental Traces}
\label{appendix:eval}

\begin{figure*}
  \begin{minipage}[b]{\columnwidth}
    \includegraphics{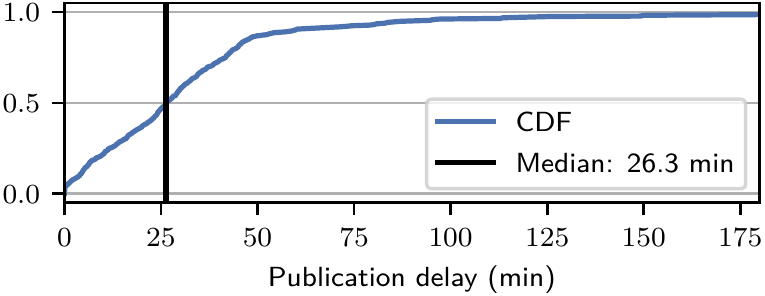}
    \caption{Reporting delays for all reports considered in \cref{sec:eval:tracking} as a cumulative distribution function.}
    \label{fig:reporting-delay}
  \end{minipage}
  \hfill
  \begin{minipage}[b]{\columnwidth}
    \centering
    \includegraphics{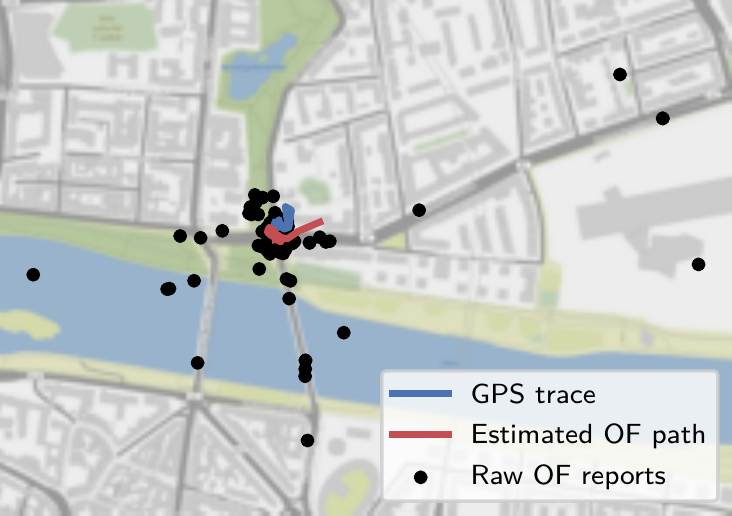}
    \caption{Map showing the GPS trace, the raw OF location reports, and the estimated paths for the \emph{restaurant} scenario (cf.~\cref{sec:eval:tracking}).}
    \label{fig:map:restaurant}
  \end{minipage}
\end{figure*}

\begin{figure*}
  \includegraphics[width=\linewidth]{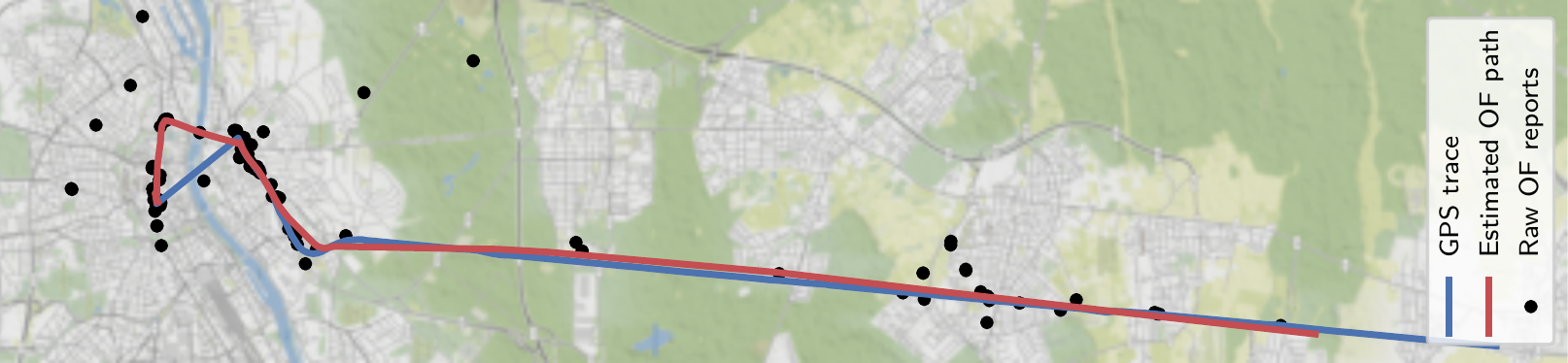}
  \caption{Map showing the GPS trace, the raw OF location reports, and the estimated paths for the \emph{train} scenario (cf.~\cref{sec:eval:tracking}).}
  \label{fig:map:train}
\end{figure*}

\begin{figure*}
  \includegraphics[width=\linewidth]{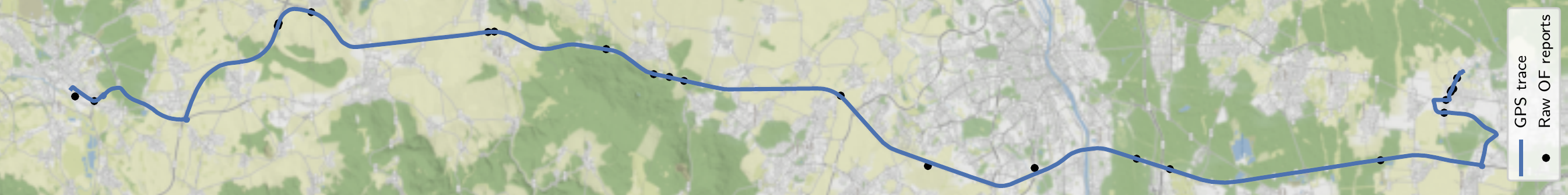}
  \caption{Map showing the GPS trace and the raw OF location reports for the \emph{car} scenario (cf.~\cref{sec:eval:tracking}).}
  \label{fig:map:car}
\end{figure*}

We show the reports of our \emph{restaurant}, \emph{train}, and \emph{car} evaluation scenarios in \cref{fig:map:restaurant,fig:map:train,fig:map:car}, respectively.

\end{document}